\pdfoutput=1
\documentclass[a4paper,11pt]{article}

\newcommand\papertitle{\boldmath 
	Symmetry-resolved entanglement for excited states and two entangling intervals in AdS${}_3$/CFT${}_2$
}

\usepackage{jheppub}
\usepackage{xcolor}
\usepackage[english]{babel}
\usepackage{tikz}
\usetikzlibrary{arrows, backgrounds, calc, decorations.pathreplacing, decorations.markings, fadings, quotes, positioning, shadings, intersections, angles}

\usepackage{amsmath,amsfonts,amssymb,mathrsfs}
\usepackage{bbm}
\usepackage[inline]{enumitem}
\usepackage{multirow}
\usepackage{subcaption}
\usepackage{nicefrac}
\usepackage{xspace}
\usepackage{mathtools}
\usepackage{graphicx} 
\graphicspath{{figures-v2/}} 
\usepackage{caption}
\usepackage{subcaption}

\usepackage[colorlinks=true]{hyperref}
\hypersetup{
	bookmarks=true,         
	unicode=false,          
	pdftoolbar=true,        
	pdfmenubar=true,        
	pdffitwindow=false,     
	pdfstartview={FitH},    
	pdftitle={\papertitle}, 
	pdfauthor={Christian Northe},
	pdfnewwindow=true,      
	colorlinks=true,        
	linkcolor=blue,         
	citecolor=red,          
	filecolor=magenta,      
	urlcolor=cyan           
}

\usepackage[colorinlistoftodos,prependcaption,textsize=tiny]{todonotes}

\makeatletter

\@addtoreset{equation}{section}
\makeatother

\newcommand{\p}	{\partial}

\newcommand{\Z}	{\mathbb{Z}}

\newcommand{\cA}{\mathcal{A}}
\newcommand{\cB}{\mathcal{B}}

\newcommand{\cH}{\mathcal{H}}

\newcommand{\cN}{\mathcal{N}}
\newcommand{\cO}{\mathcal{O}}

\newcommand{\cZ}{\mathcal{Z}}

\DeclareMathOperator{\Tr}{Tr}
\DeclareMathOperator{\tr}{\tr}

\newcommand{\bra}[1]	{\langle{#1}\vert}
\newcommand{\ket}[1]	{\vert{#1}\rangle}

\newcommand{\corr}[1]{\left\langle{#1}\right\rangle}
\newcommand{\normord}[1]{:\mathrel{#1}:}

\renewcommand{\Tr}	{\mathrm{Tr}}
\renewcommand{\tr}	{\mathrm{tr}}

\newcommand\Algebra[1]	{\mathfrak{#1}}

\renewcommand{\u}	{\hat{\Algebra{u}}}

\newcommand\ads {\text{AdS}}
\newcommand\cft {\text{CFT}}

\newcommand{\adscft}{{AdS${}_3$/CFT${}_2$}}

\newcommand{\nth}   {n^{\text{th}}}

\newcommand\Secref[1]	{Section~\ref{#1}\xspace}

\newcommand\secref[1]	{section~\ref{#1}\xspace}

\newcommand\figref[1]	{figure~\ref{#1}\xspace}
\newcommand\appref[1] {appendix~\ref{#1}\xspace}

\begin{document}
	\title{\papertitle}
	
	\author[a,1,*]{Konstantin Weisenberger,\note[1]{konstantin.weisenberger@physik.uni-wuerzburg.de}}	
	\author[a,2,*]{Suting Zhao,\note[2]{suting.zhao@physik.uni-wuerzburg.de}}
	\author[a,3]{Christian Northe,\note[3]{christian.northe@physik.uni-wuerzburg.de}}
	\author[a,4]{Ren\'e Meyer,\note[4]{Corresponding author: rene.meyer@physik.uni-wuerzburg.de
			\\
			The ordering of authors is chosen to reflect their role in the preparation of this work.}\\
		\note[*]{These authors have contributed equally to this work.}}
	\affiliation[a]{Institut f\"ur Theoretische Physik und Astrophysik\\ and \\
		W\"urzburg-Dresden Cluster of Excellence ct.qmat,\\ Julius-Maximilians-Universit\"at W\"urzburg, \\Am Hubland \\97074 W\"urzburg, Germany}
	
	\abstract{
		We test the proposal of \cite{Zhao:2020qmn} for the holographic computation of the charged moments and the resulting symmetry-resolved entanglement entropy in different excited states, as well as for two entangling intervals. Our holographic computations are performed in $U(1)$ Chern-Simons-Einstein-Hilbert gravity, and are confirmed by independent results in a conformal field theory at large central charge. In particular, we consider two classes of excited states, corresponding to charged and uncharged conical defects in AdS${}_3$. In the conformal field theory, these states are generated by the insertion of charged and uncharged heavy operators. We employ the monodromy method to calculate the ensuing four-point function between the heavy operators and the twist fields. For the two-interval case, we derive our results on the AdS and the conformal field theory side, respectively, from the generating function method of \cite{Zhao:2020qmn}, as well as the vertex operator algebra. 
		In all cases considered, we find equipartition of entanglement between the different charge sectors. 
		We also clarify an aspect of conformal field theories with a large central charge and $\u(1)_k$ Kac-Moody symmetry used in our calculations, namely the factorization of the Hilbert space into a gravitational Virasoro sector with large central charge, and a $\u(1)_k$ Kac-Moody sector. 
	}
	
	
	\maketitle
	\flushbottom

	\section{Introduction}\label{sec:Intro}

	The past decade has witnessed a fruitful interplay between quantum information theory and gravity, and in particular the physics of black holes, via the AdS/CFT correspondence \cite{Maldacena}. The AdS/CFT correspondence is a duality between strongly coupled gauge theories in $d$ space-time dimensions (the CFT side) and weakly coupled gravitational theories in $d+1$ dimensional Anti de Sitter space-times (the AdS side of the correspondence). A bridge between quantum information theory and gravity is laid by the Ryu-Takayanagi  prescription for the holographic computation of entanglement entropy in terms of the length of a minimal geodesic in the gravitational theory \cite{RT}. This triggered an avalanche of advances further translating quantum information theoretic concepts into geometric objects on the gravity side of the correspondence, for a review, c.f.  \cite{RangamaniTakayanagiBook}. Important examples include relative entropy \cite{MyersCasiniBlanco,wong2013entanglement}, quantum computational complexity \cite{Susskind:2014rva,Brown:2015bva,Brown:2015lvg}, tensor networks \cite{Swingle:2009bg,Happy},  and quantum error correction \cite{HarlowAlmheiriDong, PreskillPastawski}, entwinement \cite{Erdmenger:2019lzr}, entanglement entropy at genus one \cite{Gerbershagen:2021yma} and entanglement measures probing the entanglement shadow \cite{Gerbershagen:2021gvc}. All these examples treat CFTs at large central charge. 
	
	Most examples of quantum systems carry global symmetries.\footnote{Conservative systems have at least one global symmetry - time translation invariance.} Therefore it is interesting to investigate the entanglement associated with charged sectors in the space of states. In \cite{GoldsteinSela} this question was addressed by introducing the notion of  \textit{symmetry-resolved entanglement entropy}. It organizes the Hilbert space of a sub-interval $\cA$ into sectors of fixed charge and quantifies the entanglement for one such sectors between $\cA$ and its complement $\bar{\cA}$. Since then, this line of research has been expanded in a variety of examples including free quantum field theories \cite{Bonsignori:2019naz,Murciano:2020vgh,tan2020particle}, via dimensional reduction \cite{Murciano:2020lqq}, excited states \cite{Capizzi:2020jed}, symmetry-resolved relative entropy \cite{Chen:2021pls}, the conformal bootstrap \cite{Horvath:2021fks,Capizzi:2021kys}, entanglement negativity \cite{Murciano:2021djk}, sine-Gordon theory \cite{Horvath:2021rjd}, quenches in free fermion chains \cite{Parez:2021pgq} and Wess-Zumino-Witten models \cite{Calabrese:2021qdv}. 
	
	In this work we discuss the holographic realization of symmetry-resolved entanglement presented in \cite{Zhao:2020qmn} in charged and uncharged excited states, as well as for the case of two entangling intervals. Two main results were obtained in \cite{Zhao:2020qmn}: The first was to establish a new method based on a generating function approach to determine the symmetry-resolved entanglement entropy directly from the expectation value of the sub-region charge operator $Q_\cA$. This proved to be particularly efficient in the holographic calculation of symmetry-resolved entanglement entropy. 
	
	The second result of \cite{Zhao:2020qmn} was the proposal of a geometric dual for  symmetry-resolved entanglement entropy on the gravity side of the AdS/CFT correspondence. In \cite{Zhao:2020qmn}, we studied Einstein-Hilbert gravity in $\ads_3$ coupled to $U(1)$ Chern-Simons gauge theory, or $U(1)$ Chern-Simons-Einstein-Hilbert gravity in short. Investigating the ground state of this theory, we identified bulk Wilson lines as the natural candidates to resolve the entanglement entropy on charge sectors. The bulk Wilson line needs to align with the Ryu-Takayanagi geodesic, as required to sustain the replica symmetry of the R\'enyi entropy. Using our generating function method, we holographically derived the charged moments\footnote{In the holographic setup, the charged moments were also discussed in \cite{Belin:2013uta} where, for the vacuum state, they were related to a charged topological black hole by a  conformal transformation.} and the symmetry-resolved entanglement entropy for a CFT with $\u(1)_k$ Kac-Moody symmetry. The concrete examples studied were the case of a single entangling interval in the ground state both on the gravity and CFT side,  as well as conical defect solutions on the gravity side. Since our holographic theory carries a large central charge,  we checked our holographic result against a CFT calculation at large central charge with $U(1)$ symmetry. The key  observation in \cite{Zhao:2020qmn} was that the bulk Wilson line is dual to a charge defect in the $\cft$, which is naturally described by a pair of $U(1)$ vertex operators inserted at the endpoints of the entangling region. 
	
	While our previous work \cite{Zhao:2020qmn} focused mostly on the ground state, it is the aim of this work to test our Wilson line proposal in various excited states of the gravitational theory, and confirm each result with the corresponding CFT calculation at large central charge. As argued in \cite{Zhao:2020qmn}, the leading order contribution in entanglement is entirely fixed by replica symmetry and conformal symmetry. The examples studied in the present paper corroborate the efficiency of our generating function approach, circumventing the potentially laborious computation of the charged moments. Moreover, we compute the symmetry-resolved entanglement entropy for the case of two disconnected entanglement intervals in the ground state. While examples of the symmetry-resolved entanglement entropy for excited states can already be found in the literature for systems with small central charge, see for instance \cite{Capizzi:2020jed}, the case of holographic CFTs has only recently drawn attention \cite{Zhao:2020qmn}. The case of two entangling intervals is, to the best of our knowledge, completely unexplored so far.

	Our paper is structured as follows: \Secref{section 2} is devoted to the AdS${}_3$ side of the duality: In \Secref{section 2.1}, we first review the symmetry-resolved entanglement entropy along the lines of \cite{GoldsteinSela}, as well as the generating functional method of \cite{Zhao:2020qmn}. We focus on the main points of the calculation of the holographic charged moments in \secref{sec: holographicChargedMoment}. In \secref{section 2.3} we summarize the results of \cite{Zhao:2020qmn} for the symmetry-resolved entanglement entropy in the single interval case, and present a new result for an AdS${}_3$ background with a nonvanishing charge generated by the insertion of a pair of charged vertex operators. In \secref{section 2.4}, we present the AdS${}_3$ calculation for the case of two disjoint intervals. In \secref{section 5}, we present our CFT results. In \secref{sec:factorization}, we present arguments for the factorization of the Hilbert space of $U(1)$ Chern-Simons-Einstein-Hilbert gravity in a gravitational and a Kac-Moody sector, as well as the corresponding statement on the CFT side. In \secref{sec:UnchargedExcitation}, we use this factorization property to rederive the charged moments and symmetry-resolved entanglement entropy for an uncharged excited state. In \secref{sec:CFTChargedVacuum}, we derive corresponding results for the CFT vacuum with a background charge. In \secref{sec:CFTtwoIntervals}, we finally calculate the symmetry-resolved entanglement entropy for two disjoint intervals. We conclude and give an outlook in \secref{sec:Outlook}. Details on an alternative derivation for the case of two intervals and of the $n=2$ R\'enyi entropy using the Knizhnik-Zamolodchikov equation and the $U(1)$ Ward identity are presented in \appref{appendix B}.
	
	\section{Symmetry resolved entanglement in AdS${}_3$}\label{section 2}
	
	Given any theory with global symmetry $G$, its Hilbert space of states decomposes into irreducible representations, each corresponding to a sector of fixed charge. Then it is possible to investigate the entanglement associated with single charge sectors, i.e. the symmetry-resolved entanglement entropy. In this section, we the review previous results on the symmetry-resolved entanglement entropy of \cite{Zhao:2020qmn} and extend them to more general cases.
	

	\subsection{Symmetry resolved entanglement and the generating function method}\label{section 2.1}
	We consider a system with an internal $U(1)$ symmetry and a bipartition of its Hilbert space $\cH=\cH_\cA\otimes\cH_\cB$, associated with constant time spatial region $\cA$ and its complement $\cB$. We assume that the charge operator generating the $U(1)$ symmetry splits diagonally, $Q=Q_\cA\oplus Q_\cB$. Moreover, we restrict to eigenstates of $Q$ with density matrix $\rho$ , i.e. $[\rho,Q]=0$, implying that, after tracing out $\cB$, we have $[\rho_{\mathcal{A}},Q_{\mathcal{A}}]=0$. Therefore $\rho_{\mathcal{A}}$ is block-diagonal, with each block corresponding to an eigenvalue $q$ of the sub-region charge operator $Q_{\mathcal{A}}$, 
	\begin{eqnarray}
		\rho_{\mathcal{A}}=\oplus_{q}\rho_{\mathcal{A}}(q)\ ,\label{sum of rhoAq}
	\end{eqnarray}
	A block $\rho_\cA(q)$ is singled out by the projector $\Pi_q$ onto the eigenspace of $Q_\cA$ with fixed eigenvalue $q$,
	\begin{eqnarray}
		\rho_{\mathcal{A}}(q)=\rho_{\mathcal{A}}\Pi_{q}\ .
	\end{eqnarray}	
	Each eigenvalue $q$ is measured with probability
	\begin{eqnarray}
		P_{\mathcal{A}}(q)=\frac{\Tr\rho_{\mathcal{A}}(q)}{\Tr\rho_{\mathcal{A}}}=\Tr\rho_{\mathcal{A}}(q)\, 
	\end{eqnarray}
	to appear in a measurement of $Q_\cA$, since $\rho_\cA$ is normalized, $\Tr \rho_{\mathcal{A}}=1$. The block-diagonal structure of $\rho_{\mathcal{A}}$ enforces a block decomposition of $\rho_{\mathcal{A}}^n$,
	\begin{eqnarray}\label{sum of rhoAqn}
		\rho_{\mathcal{A}}^{n}=\oplus_{q}\rho_{\mathcal{A}}(q)^n
		\quad
		\text{with}
		\quad
		\rho_{\mathcal{A}}(q)^n=\left(\rho_{\mathcal{A}} \Pi_{q}\right)^n=\rho_{\mathcal{A}}^n\Pi_q\,,
	\end{eqnarray}
	and the probability distribution of a charge block in $\rho_{\mathcal{A}}^n$ is given by
	\begin{eqnarray}
		P_{\mathcal{A},n}(q)=\frac{\Tr\rho_{\mathcal{A}}(q)^n}{\Tr\rho_{\mathcal{A}}^n}\,.\label{Pn}
	\end{eqnarray}
	The decompositions \eqref{sum of rhoAq} and \eqref{sum of rhoAqn} can be used to define the \textit{symmetry-resolved R\'enyi entropy},
	\begin{eqnarray}
		S_{n}(q)=\frac{1}{1-n}\log \Tr\left(\frac{\rho_{\mathcal{A}}(q)}{P_{\mathcal{A}}(q)}\right)^{n}=\frac{1}{1-n}\log{\frac{\mathcal{Z}_{n}(q)}{\mathcal{Z}_{1}(q)^{n}}}\ , \label{SRRenyi}
	\end{eqnarray}
	where we denote
	\begin{eqnarray}
		\mathcal{Z}_{n}(q)=\Tr\rho_{\mathcal{A}}(q)^n=\Tr \left[\rho_{\mathcal{A}}^n\Pi_q\right]\ ,\label{Znq}
	\end{eqnarray}
	The entropies \eqref{SRRenyi} are measures of the amount of entanglement between the subsystems $\mathcal{A}$ and $\mathcal{B}$ in each charged block.
	
	The \textit{symmetry-resolved entanglement entropy} in the $q$-sector  is given by the $n\to 1$ limit of $S_{n}(q)$,
	\begin{eqnarray}
		S_{1}(q)=\lim_{n\to 1} S_{n}(q)=-\Tr\left[ \frac{\rho_{\mathcal{A}}(q)}{P_{\mathcal{A}}(q)}\log\left(\frac{\rho_{\mathcal{A}}(q)}{P_{\mathcal{A}}(q)}\right)\right]\ . \label{S1q}
	\end{eqnarray}
	From \eqref{sum of rhoAq} and \eqref{S1q}, the total entanglement entropy associated with $\rho_{\mathcal{A}}$ decomposes into
	\begin{eqnarray}
		S_{1}=\sum_{q}P_{\mathcal{A}}(q)S_{1}(q)-\sum_{q}P_{\mathcal{A}}(q)\log P_{\mathcal{A}}(q)\ .\label{S1-S1q}
	\end{eqnarray}
	The two terms in \eqref{S1-S1q} are usually referred to as \textit{configurational} and \textit{fluctuation} entropy, respectively \cite{Lukin256}, the former being related to the operationally accessible entanglement entropy \cite{WisemanVaccaro, BarghathiHerdman, BarghathiCasiano}.
	
	As can be seen from \eqref{SRRenyi}, calculating $\mathcal{Z}_{n}(q)$ is necessary to obtain the symmetry-resolved R\'enyi and entanglement entropies. However, this is generally very involved as it demands knowledge of the spectrum of the reduced density matrix $\rho_{\mathcal{A}}$ and its resolution in $Q_{\mathcal{A}}$.
	Therefore, the idea advocated in \cite{GoldsteinSela} is rather to compute the \textit{charged moments}
	\begin{eqnarray}\label{chargedMoments}
		\mathcal{Z}_{n}(\mu)=\Tr\left[\rho_{\mathcal{A}}^{n}e^{i\mu Q_{\mathcal{A}}}\right]\,.
	\end{eqnarray}
	These are related to \eqref{Znq} by Fourier transformation,
	\begin{eqnarray}\label{chargedMomentsFourier}
		\mathcal{Z}_{n}(q)=\int_{-\infty}^{\infty}\frac{d\mu}{2\pi} e^{-i\mu q}\mathcal{Z}_{n}(\mu)\,,
	\end{eqnarray}
	for the case of $q\in \mathbbm{R}$. For discrete eigenvalues, for example if $\mathcal{Z}_{n}(\mu)=\mathcal{Z}_{n}(\mu+2\pi)$, the range of integration \eqref{chargedMomentsFourier} changes to $[-\pi,\pi]$.
	
	While the charged moments \eqref{chargedMoments} provide a way to calculate the symmetry-resolved entropies, they are not easily computed for general excited states (c.f. \cite{Capizzi:2020jed, Horvath:2020vzs, Bonsignori:2020laa}). 
	Based on a generating function, a new method which simplifies the calculation was introduced in \cite{Zhao:2020qmn}. It extracts the charged moments and hence the symmetry-resolved entanglement entropy directly from the expectation value of the sub-region charge operator $Q_\cA$, which is generally easier to compute than the charged moments themselves. Here we recapitulate this method and in subsequent sections we provide examples corroborating its efficiency in the context of holography.
	
	Define a normalized generating function associated with the charged moments $\mathcal{Z}_{n}(\mu)$ as
	\begin{eqnarray}\label{generatingfunction}
		f_{n}(\mu)\coloneqq \frac{\mathcal{Z}_{n}(\mu)}{\mathcal{Z}_{n}(0)}=\frac{\mathcal{Z}_{n}(\mu)}{\mathcal{Z}_{n}},
	\end{eqnarray}
	fulfilling the initial condition 
	\begin{equation}\label{initialCondition}
		f_{n}(0)=1\,.
	\end{equation}
	The expectation value of $Q_A$ is then expressed as
	\begin{eqnarray}\label{chargeExpectationValue}
		\langle iQ_{\mathcal{A}}\rangle_{n,\mu}
		:=
		\frac{\Tr \left[iQ_{\mathcal{A}} \rho^{n}e^{i\mu Q_{\mathcal{A}}}\right]}{\Tr\left[\rho^{n}e^{i\mu Q_{\mathcal{A}}}\right]}
		=
		\frac{\partial \ln \cZ_{n}(\mu)}{\partial\mu}
		=
		\frac{\partial \ln f_{n}(\mu)}{\partial \mu}\ .\label{Q-f}
	\end{eqnarray}
	In order to evaluate $f_n(\mu)$, and hence $Z_n(\mu)$, it suffices to know the expectation value of the charge, and integrate \eqref{chargeExpectationValue} with initial condition \eqref{initialCondition}.
	
	As evident from \eqref{Pn}, \eqref{Znq} and \eqref{chargedMomentsFourier}, the probability distribution $P_{\mathcal{A},n}(q)$ can be recast in terms of the generating function by a Fourier transformation, 
	\begin{eqnarray}\label{probabilityGeneratingFunction}
		P_{\mathcal{A},n}(q)=\frac{\mathcal{Z}_n(q)}{\mathcal{Z}_n}=\int_{-\infty}^{\infty}\frac{d\mu}{2\pi} e^{-i\mu q}f_{n}(\mu)\ ,
	\end{eqnarray}
	As pointed out in \cite{Capizzi:2020jed}, the symmetry-resolved R\'enyi entropy,
	\begin{eqnarray}
		S_{n}(q)=S_n+\frac{1}{1-n}\log{\frac{P_{\mathcal{A},n}(q)}{P_{\mathcal{A}}(q)^{n}}}\ ,\label{SRE simp}
	\end{eqnarray}
	as well as the symmetry-resolved entanglement entropy
	\begin{eqnarray}
		S(q)=S+\lim_{n\to 1}\frac{1}{1-n}\log{\frac{P_{\mathcal{A},n}(q)}{P_{\mathcal{A}}(q)^{n}}}\ ,\label{SEE simp}
	\end{eqnarray}
	split into two pieces. The first are their uncharged counterparts, the R\'enyi entropy $S_n$ and the entanglement entropy $S$, while the charge information resides fully in the second contribution.

	\subsection{Holographic charged moments}\label{sec: holographicChargedMoment}
	The holographic dual of charged moments in $U(1)$ Chern-Simons-Einstein gravity was first discussed in \cite{Belin:2013uta}. Here we follow the Wilson line defect prescription of \cite{Zhao:2020qmn}, and briefly review the holographic interpretation of the charged moments.\\
	
	In $U(1)$ Chern-Simons-Einstein-Hilbert gravity, two additional chiral gauge fields are introduced on top of the gravitational sector. The asymptotic behavior of the left-moving gauge field takes the following form in Fefferman-Graham coordinates \cite{Kraus:2006wn},
	\begin{align}
		A=A^{(0)}+e^{-2\rho}A^{(2)}+\cdots\ \quad \textrm{as}\quad \rho\to \infty\ ,
	\end{align}
	where $\rho$ is the radial coordinate in $\ads$ and, for convenience, we have set the $\ads$ radius $l=1$. Under this asymptotic expansion and employing the boundary condition $A^{(0)}_{\rho}=0$, the variation of the action with respect to $A$ yields \cite{Kraus:2006wn}
	\begin{align}
		\delta I_{A}=\frac{i}{2\pi}\int_{\partial \mathcal{M}} d^2x \sqrt{g^{(0)}}J^a \delta A_{a}^{(0)}\,.
	\end{align}
	Here, $J^{a}$ is identified as the boundary $U(1)$ current whose components in complex coordinates read
	\begin{align}\label{current def}
		J_{w}=\frac{1}{2}J^{\bar{w}}=\frac{i k}{2}A_{w}^{(0)}\ ,\quad J_{\bar{w}}=0\ ,
	\end{align}
	Since $J_{\bar{w}}$ always vanishes, the corresponding source term is $A^{(0)}_{\bar{w}}$. By the AdS/CFT correspondence, in the presence of a source term, the CFT action is deformed as \cite{Kraus:2006nb,Kraus:2006wn}
	\begin{align}
		I_{CFT}\to I_{CFT}+\frac{i}{2\pi} \int_{\partial \mathcal{M}}d^2w J^{\bar{w}} A^{(0)}_{\bar{w}}\ .
	\end{align}
	The left-moving sub-region charge on $\mathcal{A}$ is defined as
	\begin{align}\label{sub-region charge def}
		q_{\mathcal{A}}=\int_{\mathcal{A}}\frac{dw}{2\pi i} J_{w}\ .
	\end{align}
	The idea relating the charged moments with the deformation of the CFT action sets out by rewriting the sub-region charge \eqref{sub-region charge def} as a two-dimensional integral over the boundary Riemann surface $\partial \mathcal{M}$. For instance, considering $\mathcal{A}$ to be a single interval and denoting the endpoints of the interval $\mathcal{A}$ as $w=z_1$ and $w=z_{2}$, one can rewrite \eqref{sub-region charge def} as 
	\begin{align}
		q_{\mathcal{A}}=\int_{z_2}^{z_1}\frac{dw}{2\pi i}J_{w}=\int_{\partial \mathcal{M}}\frac{d^2w}{2\pi i}\left(G(\bar{w},\bar{z}_1)-G(\bar{w},\bar{z}_2)\right) J_{w}\ ,
	\end{align}
	where $G(\bar{w},\bar{z})$ is the Green kernel of the $\partial$ operator defined on the boundary Riemann surface. Therefore, the additional operator in the charged moments can be thought of as introducing an additional source term coupled to the $U(1)$ current, where the additional source reads
	\begin{align}\label{source 1}
		\delta A^{(0)}_{\bar{w}}=i\mu \left(G(\bar{w},\bar{z}_1)-G(\bar{w},\bar{z}_2)\right)\ .
	\end{align}
	
	As shown in \cite{Zhao:2020qmn}, this kind of source is realized in the bulk by introducing a $U(1)$ Wilson line defect action 
	\begin{align}\label{U(1) defect}
		I_{d}=-\frac{ik\mu}{2\pi}\int_{\mathcal{C}}A\,,
	\end{align}
	where the curve $\mathcal{C}$ is anchored at the endpoints of the boundary interval $\mathcal{A}$ and is required to be orthogonal to the boundary when approaching spatial infinity. Observe that the $U(1)$ defect action is actually topological in the bulk.
	
	In principal, different choices of the curve $\mathcal{C}$ result in different gauge field configurations in the bulk. As argued in \cite{Zhao:2020qmn} a natural choice of the curve $\mathcal{C}$ is the geodesic anchored at the endpoints of $\mathcal{A}$.  It was shown in \cite{Zhao:2020qmn} that approaching spatial infinity, independent of the shape of the Wilson line, the field strength always takes the form 
	\begin{align}
		\lim_{\rho\to\infty}F&=F_{w\bar{w}}dw\wedge d\bar{w}\nonumber\\
		&=2\mu\left(\delta^{(2)}(w-z_1,\bar{w}-\bar{z}_1)-\delta^{(2)}(w-z_2,\bar{w}-\bar{z}_2)\right)dw\wedge d\bar{w}\,.
	\end{align}
	Hence, the leading order in the asymptotic expansion of the gauge field $A$ is fixed as
	\begin{align}\label{full connection}
		A^{(0)}_{w}&=a(w)-i\mu \left(G(w,z_1)-G(w,z_2)\right)\,,\nonumber\\
		A^{(0)}_{\bar{w}}&=\bar{a}(\bar{w})+i\mu \left(G(\bar{w},\bar{z}_1)-G(\bar{w},\bar{z}_2)\right)\,.
	\end{align}
	Here, $a(w)$ and $\bar{a}(\bar{w})$ denote general background connections. From this the current \eqref{current def} is read out and the sub-region charge as well as the generating function $f_{n}(\mu)$ can be computed.
	
	\subsection{Symmetry resolved entanglement for a single interval}\label{section 2.3}
	In this section we holographically derive the symmetry-resolved entanglement entropy for excited states and a single entangling interval. The contribution from the Wilson line defect to the charged moments and the generating function $f_{n}(\mu)$ is determined by the complex structure of the boundary replica $n$-fold, without being affected by the details of the bulk geometry. Furthermore, we show that the probability distributions $P_{\mathcal{A},n}(q)$ are the Gaussian distributions of the charge fluctuations in the general case. This demonstrates that the symmetry-resolved entanglement entropy is independent of the sub-region charge in all these cases for holographic $U(1)$ Chern-Simons theory.	
	
	For Poincar\'e $AdS_3$, we denote the boundary complex coordinates as $z$ and $\bar{z}$. Without loss of generality, we consider a background gauge field $A$ with general asymptotic expansion
	\begin{eqnarray}\label{bg connection}
		A_{z}=a(z)+\cdots\ ,\quad \quad A_{\bar{z}}=\bar{a}(\bar{z})+\cdots \quad\textrm{as}\quad \rho\to \infty
	\end{eqnarray}
	where the leading order terms in \eqref{bg connection} imply a non-vanishing background $U(1)$ current. For a single boundary interval with endpoints located at $z=z_1$ and $z=z_2$, the charged moments $\mathcal{Z}_{n}(\mu)$ are constructed by inserting the Wilson line defect \eqref{U(1) defect} in the bulk repica $n$-fold. Since we are interested in the sub-region charge giving rise to the charged moments $\cZ_n(\mu)$, in order to employ \eqref{full connection}, we need the Green kernel of $\p_{\bar{z}}$ on the replica $n$-fold. This can be derived from the Green kernel in the complex plane,
	\begin{align}
		G(\bar{z},\bar{z}_i)=\frac{1}{2\pi}\frac{1}{\bar{z}-\bar{z}_i}\ , \quad i=1,2 \ ,
	\end{align}
	as follows.
	The topology of the replica $n$-fold is still a Riemann sphere in the single interval case. The replica $n$-fold is mapped to a single complex $\xi$-plane by the uniformization mapping
	\begin{align}\label{uniformization}
		\xi=\left(\frac{z-z_2}{z-z_1}\right)^{\frac{1}{n}}\,.
	\end{align}
	A fixed $z$ corresponds to $n$ different values of $\xi$. The mapping \eqref{uniformization} can be used to obtain the Green kernel of the replica $n$-fold in $z$-coordinates as\footnote{Since, in this case, the topology is trivial, the Green kernel transforms as an anti-holomorphic one-form.}
	\begin{align}\label{eq:GreenKernel1}
		G^{(n)}(\bar{z},\bar{z}_i)=\left(\frac{d\xi}{dz}\right) G(\bar{\xi},\bar{\xi}_i)=\frac{1}{2\pi n}\frac{1}{\bar{z}-\bar{z}_i}\ , \quad i=1,2 \ .
	\end{align}
	It is this Green kernel which, when plugged into \eqref{full connection}, provides the gauge field configuration for the $\cZ_n(\mu)$ charged moments.

	The next task is to obtain the background gauge field in the replica $n$-fold without additional $U(1)$ Wilson line defect insertion. As we show below, the leading order of the Chern-Simons connection in the replica $n$-fold always takes the same form as the background solution in \eqref{bg connection}. The differences between them are hidden in the $n$-sheeted coordinates $z$ and $\bar{z}$. In fact, the connection in the replica $n$-fold is an analytic continuation of the background connection from the original manifold.

	Indeed, consider a background gauge field $a(z)$ in \eqref{bg connection} generated by the boundary operator excitations located at $z=0$ and $z=\infty$. 
	\begin{figure}[t]
		\centering
		\includegraphics[width=0.5\linewidth]{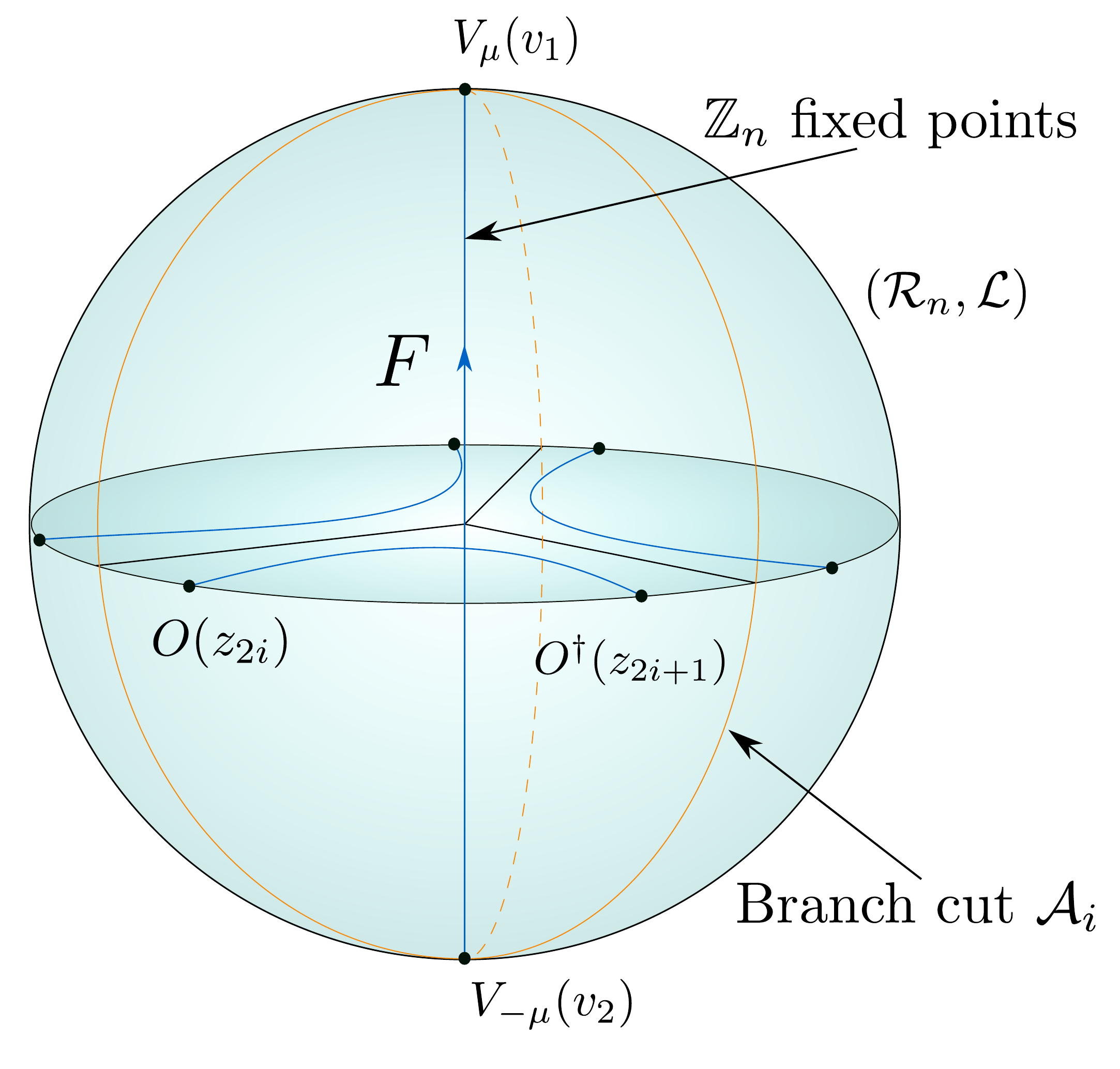} 
		\caption{ 
			For a single entangling interval, the universal covering of the replica $n$-fold is a Riemann sphere. There are $n$ copies of excitations on the replica $n$-fold. The blue lines represent the bulk Wilson lines dual to the boundary excitations. The orange lines represent the branch cuts between the $n$ sheets in the original replica manifold.}\label{fig:replicaExcited}
	\end{figure}
	Due to this pole structure, the  connection can be written as $a(z)=\frac{\alpha}{z}$, with $\alpha$ some constant associated with the charge of the excited operators. As shown in \figref{fig:replicaExcited}, in the replica trick, one inserts $n$ copies of the excited operators in the replica $n$-fold, which are located on each sheet at $z_{2m-1}=0$ and $z_{2m}=\infty$ $(m=1,\cdots, n)$, with $m$ denoting the $m^{\text{th}}$ sheet of the replica $n$-fold.
	By the uniformization mapping \eqref{uniformization}, the excited operators are mapped to the new $2n$ points $\xi_{2m-1}=\left(\frac{z_2}{z_1}\right)^{\frac{1}{n}}e^{2\pi i m /n}$ and $\xi_{2m}=e^{2\pi i m/n}$ $(m=1,\cdots, n)$. Therefore, in the complex $\xi$-plane, one can write the leading order background connection in the replica $n$-fold as
	\begin{align}\label{n connection}
		A^{(0)}_\xi=\sum_{m=1}^{n}\left(\frac{\alpha}{\xi-\xi_{2m-1}}-\frac{\alpha}{\xi-\xi_{2m}}\right)\,.
	\end{align}
	The connection \eqref{n connection} on the replica $n$-fold  is related to the original connection $a(z)=\frac{\alpha}{z}$ by a conformal transformation, hence
	\begin{align}
		A^{(0)}_{\xi}\left(\frac{d\xi}{dz}\right)=a(z)\,.
	\end{align}
	In the $n$-sheeted $z$-coordinates, the background connection of the replica $n$-fold hence still takes the same form as the original connection $a(z)$.
	
	We are now ready to derive the sub-region charge with an additional Wilson line defect insertion. From \eqref{sub-region charge def} together with \eqref{current def} it follows that
	\begin{align}
		q_{\mathcal{A}}&=\frac{ik}{2}\int_{z_2+\delta}^{z_1-\delta}\frac{dz}{2\pi i}\left(a(z)-i\mu G^{(n)}(z,z_1)+i\mu G^{(n)}(z,z_2)\right)\nonumber\\
		&=q_0+\frac{ik\mu}{4\pi}\log{\left(\frac{z_1-z_2}{\delta}\right)}\,.
	\end{align}
	Here $q_0=\frac{ik}{2}\int_{z_2}^{z_1}\frac{dz}{2\pi i}\ a(z)$ is the left-moving background sub-region charge, which is finite. The right-moving part takes an analogous form. The definition of the generating function \eqref{chargeExpectationValue} then implies
	\begin{align}
		f_{n}(\mu)&=\exp{\left(i\int_{0}^{\mu} d\mu'\ \langle \hat{Q}_{\mathcal{A}}\rangle_{n,\mu'}\right)} \label{generatingPP}
		=
		e^{i\mu Q_0}\cdot\left|\frac{v_1-v_2}{\delta}\right|^{-\frac{k}{n}(\frac{\mu}{2\pi})^2}\,,
	\end{align}
	where $Q_0=q_0+\bar{q}_0$ is the total sub-region charge originating purely from the background solutions \eqref{bg connection}.
	Using the regularized length of the geodesic in Poincar\'e AdS,
	\begin{eqnarray}
		L=2\ln{\left|\frac{v_1-v_2}{\delta}\right|}\ ,
	\end{eqnarray}
	where $\delta$ is related to the bulk UV-cutoff $\rho_0$ via $\delta=e^{-\rho_0}$, and is the cutoff around the boundary endpoints of $\cA$. The sub-region charge and the normalized generating function are expressed through
	\begin{eqnarray}
		\langle \hat{Q}_{\mathcal{A}}\rangle_{n,\mu}=Q_0+\frac{i k\mu}{4\pi^2 n}L\ ,\quad f_{n}(\mu)=e^{i\mu Q_0-\frac{k}{2n}(\frac{\mu}{2\pi})^2 L}\ . \label{ex 1 Q f}
	\end{eqnarray}
	The Fourier transformation of $f_{n}(\mu)$ in \eqref{ex 1 Q f} yields the probability distribution $P_{\mathcal{A},n}(q)$
	\begin{eqnarray}\label{Pdistribution}
		P_{\mathcal{A},n}(q)=\int_{-\infty}^{\infty}\frac{d\mu}{2\pi} e^{-i\mu q}f_{n}(\mu)=\sqrt{\frac{2\pi n}{k L}}e^{-\frac{2n\pi^2\Delta q^2}{k L}}\,,
	\end{eqnarray}
	where the fluctuation of the sub-region charge is denoted by $\Delta q=q-Q_0$.
	The resulting symmetry-resolved entanglement entropy \eqref{SEE simp} is then easily obtained,
	\begin{eqnarray}\label{chargedAdSsymmetry-resolved entanglement entropy}
		S(q)=S+\lim_{n\to 1}\frac{1}{1-n}\log{\frac{P_{\mathcal{A},n}(q)}{P_{\mathcal{A}}(q)^n}}=\frac{c}{6}L-\frac{1}{2}\ln{\left(\frac{kL}{2\pi}\right)}+O(1)\ ,
	\end{eqnarray}
	where the entanglement entropy $S$ in three dimensions is given by the Ryu-Takayanagi formula \cite{RT} $S=\frac{c}{6}L$. In \eqref{chargedAdSsymmetry-resolved entanglement entropy}, the charge dependence has disappeared in the final result, i.e. there is equipartition of entanglement amongst the charge sectors \cite{Xavier:2018kqb}.
	
	For global AdS or conical defect solutions, one obtains the symmetry-resolved entanglement entropy straightforwardly from the above procedure as well. The main difference to the case of Poincar\'e AdS, c.f. \eqref{eq:GreenKernel1}, is that the boundary Green kernel now takes the form 
	\begin{equation}
		G^{(n)}(w,w_i) = \frac{1}{2\pi n} \frac{e^{i w_i}}{e^{i w} - e^{i w_i}}\,, \quad w = \phi + i t_E\,.
	\end{equation}	
	Here $w$ are Euclidean cylinder coordinates. As shown in \cite{Zhao:2020qmn}, the final result for the symmetry-resolved entanglement entropy in the case of global AdS or conical defects is given by
	\begin{eqnarray}
		S(q)=S-\frac{1}{2}\log{\left(\frac{kL_1}{2\pi}\right)}+O(1)\ ,\quad 
		L_1 =  2 \log \left| \frac{2}{\delta} \sin \frac{\Delta\phi}{2}\right|\,, 
		\label{eq:symmetry-resolved entanglement entropyConicalDefect}
	\end{eqnarray}
	where $\Delta\phi$ is the angular extend of the boundary interval, and $S$ is the usual entanglement entropy contribution. An interesting point is that the contribution from the $U(1)$ sector in \eqref{eq:symmetry-resolved entanglement entropyConicalDefect} is not sensitive to the existence of the conical defects at all. In other words, the contribution from the $U(1)$ sector is completely topological, and only senses the complex structure of the boundary Riemann surface. This is to be expected from the topological nature of  the bulk Chern-Simons theory, whose only propagating degrees of freedom are boundary photons \cite{belin2020bulk}. Again, the equipartition still holds in all those cases.
	
	\subsection{Symmetry resolved entanglement for two disjoint intervals}\label{section 2.4}
	

	\begin{figure}
	\begin{subfigure}{1\textwidth}
		\centering
		\includegraphics[width=0.8\linewidth]{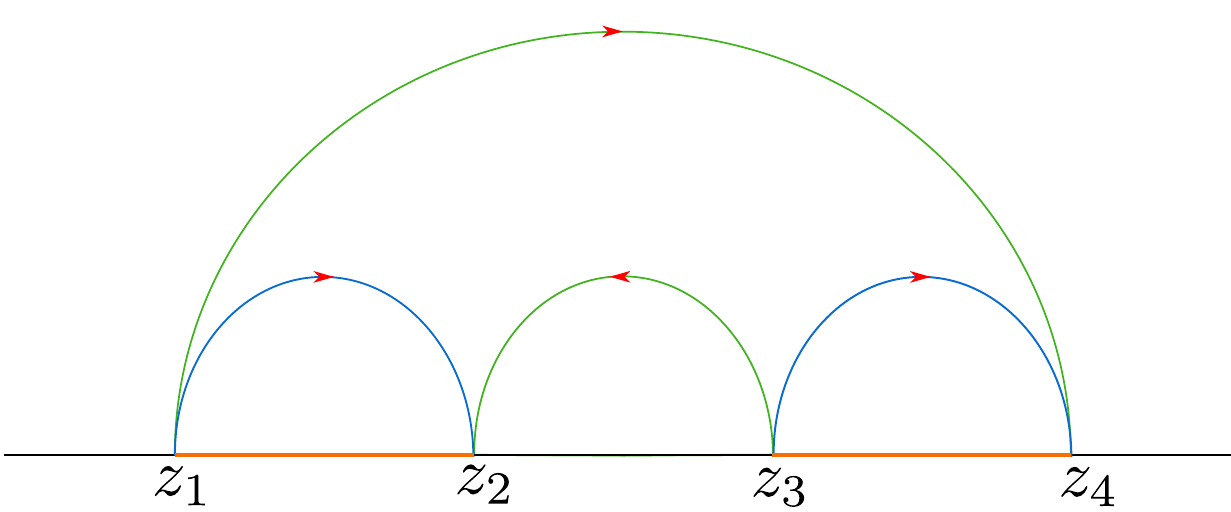}
		\caption{}
	\end{subfigure}%
\\[1cm]
    \begin{subfigure}{0.5\textwidth}
        \centering
		\includegraphics[width=0.8\textwidth]{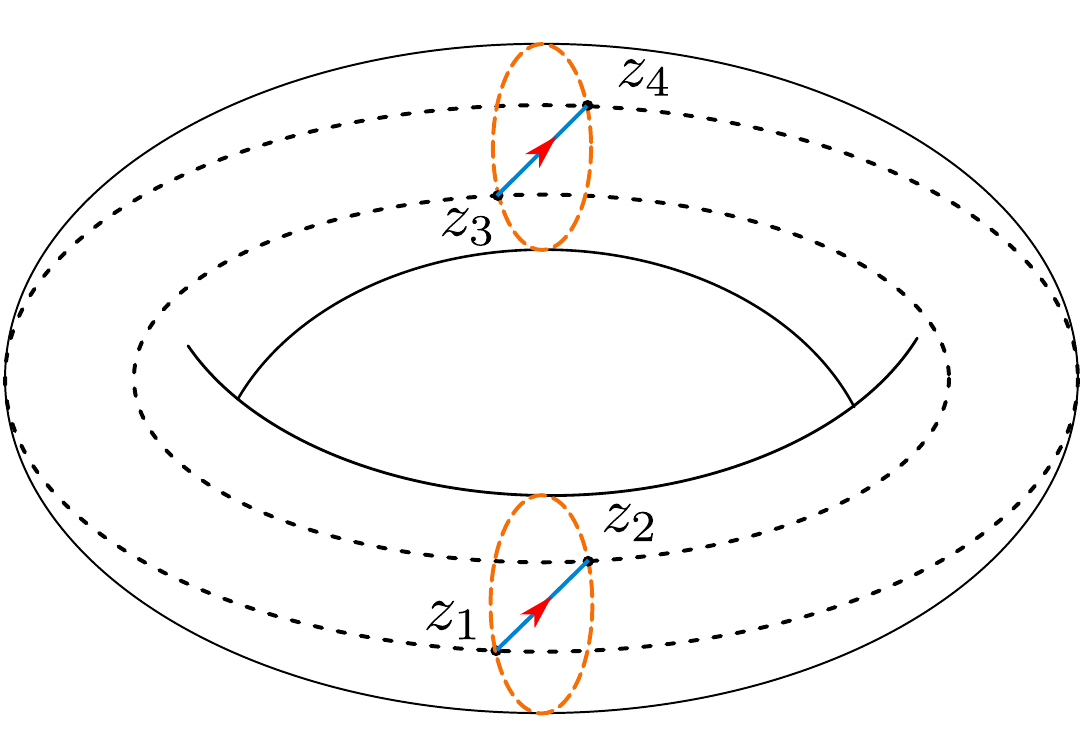}
		\caption{}
    \end{subfigure}%
		\hspace{0.1truecm}%
	\begin{subfigure}{0.5\textwidth}
        \centering
		\includegraphics[width=0.8\textwidth]{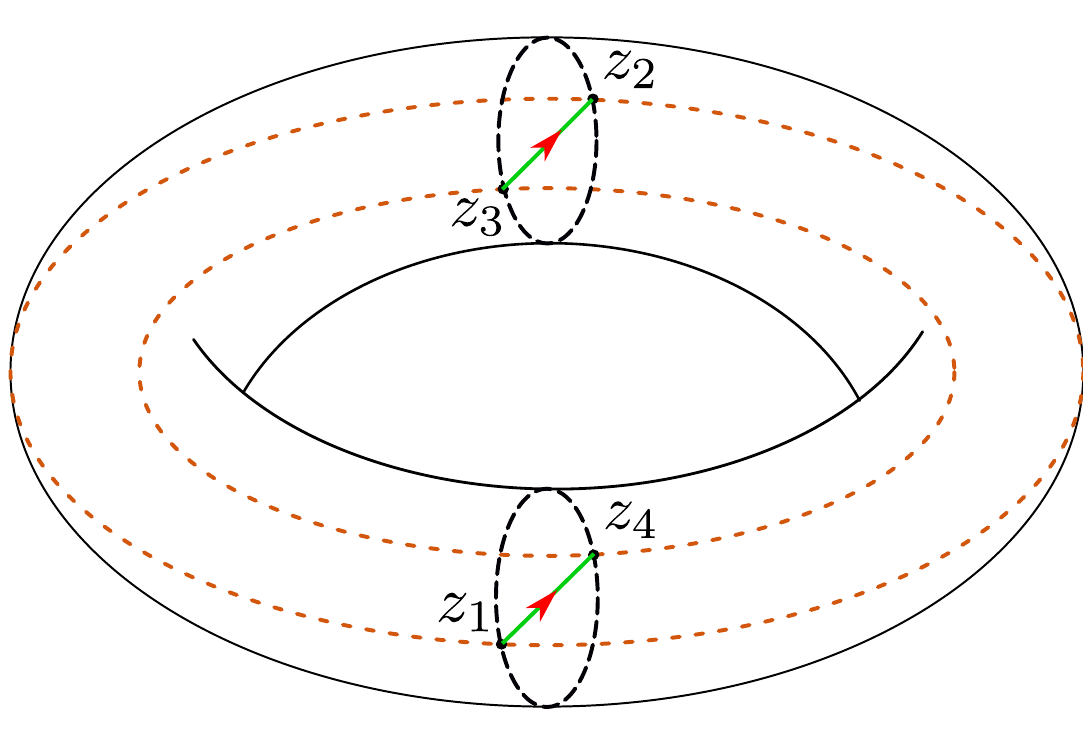}  
		\caption{}
	\end{subfigure}	
		\caption{(a) Shows the two different configurations of the Wilson lines in the bulk replica $n$-fold, where the blue and green lines are the Wilson lines dual to the $s$-channel and $t$-channel of the four point function of charged twist operators, respectively. The red arrows indicate the direction of the field strength. (b) and (c) show the bulk constructions in the universal covering of the replica $2$-fold corresponding to $s$-channel and $t$-channel, respectively. In each phase, the Wilson lines always follow the locus of the $\Z_2$ fixed points of the torus, which are the geodesics with minimal length. The branch cuts are represented by the orange lines, which are contractible circles in the $s$-channel and non-contractible in the $t$-channel.}
		\label{fig:Torus}
	\end{figure}
	
	In this section, we consider the case of two disjoint intervals. Following \cite{Zhao:2020qmn}, for an entangling region $A=A_1\cup A_2$ consisting of two disjoint intervals $A_1\cap A_2=\varnothing$ with endpoints located at $z=z_i$ $(i=1,2,3,4)$, the holographic dual of the CFT charged moments is given by inserting two $U(1)$ Wilson lines along the locus of the $\mathbb{Z}_n$ replica symmetry fixed points in the bulk. Due to the phase transition in the R\'enyi entropy in the two interval case, there exist two different bulk Wilson line defect configurations, as we shown in \figref{fig:Torus}. Although the bulk constructions are different, we show that the contribution to the charged moments from the $U(1)$ sector are identical. Again, we find equipartition behavior.
	
	We consider Poincar\'e AdS with vanishing background gauge field. The calculation for the sub-region charge is still straightforward. Analogous to \eqref{source 1}, in the two  interval case, the charged moments $\mathcal{Z}_{n}(\mu)$ require the introduction of a source term, 
	\begin{align}\label{243}
		A^{(0)}_{\bar{z}}=\frac{i\mu}{2\pi n} \sum_{m=1}^{2}\left(\frac{1}{\bar{z}-\bar{z}_{2m-1}}-\frac{1}{\bar{z}-\bar{z}_{2m}}\right)\,.
	\end{align}
	The corresponding holomorphic part of the connection reads
	\begin{align}
		A^{(0)}_{z}=-\frac{i\mu}{2\pi n} \sum_{m=1}^{2}\left(\frac{1}{z-z_{2m-1}}-\frac{1}{z-z_{2m}}\right)\,.
	\end{align}
	Evaluating the sub-region charge yields the generating function
	\begin{align}
		f_{n}(\mu)=\left|\frac{(z_1-z_2)(z_3-z_4)(z_1-z_4)(z_2-z_3)}{\delta^2(z_1-z_3)(z_2-z_4)}\right|^{-\frac{k\mu^2}{4\pi^2 n}}
	\end{align}
	The symmetry-resolved entanglement entropy is given by
	\begin{align}\label{Sqtwointervalsgravity}
		S(q)=S-\frac{1}{2}\log\left(\frac{k}{\pi}\log\left|\frac{(z_1-z_2)(z_3-z_4)(z_1-z_4)(z_2-z_3)}{\delta^2(z_1-z_3)(z_2-z_4)}\right|\right)\,.
	\end{align}
	This rather simple calculation contains a nontrival bulk interpretation for the Wilson line defect configurations. Observe that the phase transition for symmetry-resolved entanglement entropy still takes place at the usual critical point, i.e.  $x=\frac{(z_1-z_2)(z_3-z_4)}{(z_1-z_3)(z_2-z_4)}=\frac{1}{2}$. It arises from different dominating contributions in the $s$- and $t$-channel of the four point function of twist operators \cite{Hartman:2013mia}. On the CFT side, these two situations are represented by imposing trivial monodromy conditions for different cycles \cite{Hartman:2013mia} around points $(z_1,z_2)$ or $(z_2,z_3)$, respectively. 
	
	On the gravity side, these two different monodromy conditions indicate which circle of the bulk geometry is contractible \cite{faulkner2013}. 
	Both bulk constructions should preserve the $\mathbb{Z}_n$ symmetry. What distinguishes the two phases is the locus of the bulk $\mathbb{Z}_n$ fixed points. Since the $U(1)$ Wilson lines are required to respect the $\mathbb{Z}_n$ symmetry in the two different phases, they follow different paths in the replica manifold, i.e. the respective $Z_n$ fixed points. Nevertheless, the boundary value of the gauge field takes the same form \eqref{243} in $z$-coordinates. As a result, the phase transition in the symmetry-resolved entanglement entropy is not affected by the $U(1)$ sector, and happens at the same cross ratio as for entanglement and R\'enyi entropies.
	
	\section{Symmetry resolved entanglement in CFT${}_2$}\label{section 5}
	In this section we confirm our holographic calculations with CFT results at large central charge. We begin in \secref{sec:factorization} with an analysis of the space of states of the CFT, which was implicitly employed in \cite{Zhao:2020qmn}. Namely, we argue that the total Hilbert space $\cH_{tot}$ factorizes into a gravity sector $\mathcal{H}_{g}$ and a $\u(1)_k$ Kac-Moody sector $\mathcal{H}_{k}$. This paves the way to first reproduce the vacuum charged moments, after which we present CFT calculations of charged moments and symmetry-resolved entanglement entropy for the holographic dual of the conical defect solution in \secref{sec:UnchargedExcitation}, charged Poincaré AdS in \secref{sec:CFTChargedVacuum} and investigate the two interval case in \secref{sec:CFTtwoIntervals}.
	
	\subsection{Factorization of the Hilbert space}\label{sec:factorization}
	We now investigate the structure of the Hilbert space of the boundary theory by taking a close look at bulk theory. Since we are working in the bottom-up set-up of the AdS/CFT correspondence, the Hilbert space for the bulk (or boundary) theory is only a subspace of the full Hilbert space in a putative full UV complete top-down approach. In particular, the decoupling of the bulk $U(1)$ Chern-Simons field from the bulk metric implies that one can change the configuration of the $U(1)$ gauge field without affecting the bulk geometry. This indicates that the bulk states considered here are factorized into a gravitational part and a $U(1)$ part. We exclude situations in which this factorization will not hold, for example in the presence of additional charged bulk fields which are coupled to both the metric and the $U(1)$ gauge field \cite{KrausPrivate}, or bottom-up models with a different symmetry structure such as e.g. Einstein-Maxwell \cite{perez2016asymptotic}. 
	
	Evidence for this factorization property can be traced back to the asymptotic symmetry algebra of $U(1)$ Chern-Simons-Einstein-Hilbert theory, which is the starting point of our CFT analysis. As shown in \cite{Zhao:2020qmn}, the modes of the stress tensor can be expressed as $L_n=L_n^g+L_n^k$, where $L_n^g$ and $L_n^k$ are the Virasoro modes in the gravity and the $\u(1)_k$ sector, respectively. From the bulk decoupling of the metric and $\u(1)_k$ sector it follows that $[L_n^g,L_m^k]=0$ and $[L_n^g,J_m]=0$. This split stress-energy tensor reproduces the correct Kac-Moody algebra
	\begin{align}\label{symmalgebra}
		[J_{n}, J_m]&=\frac{1}{2}n k \delta_{m+n}\ ,\nonumber\\
		[L_{n}, J_m]&= -m J_{n+m}\ ,\nonumber\\
		[L_{n}, L_m]&=(n-m)L_{n+m}+\frac{c}{12}(n^3-n)\delta_{n+m,0}\ ,
	\end{align}
	where the central charge is the Brown-Henneaux central charge $c_g=\frac{3l}{2G_3}$ shifted by the central charge of the $\u_1(k)$ theory $c_k=1$, i.e. $c=c_g+c_k$. A detailed analysis for the effective central charge of the $U(1)$ Chern-Simons Einstein gravity is provided in \cite{belin2020bulk}. 
	Due to the vanishing commutators $[L_n^g,L_m^k]=0$ and $[L_n^g,J_m]=0$, \eqref{symmalgebra} decomposes into a gravitational Virasoro algebra for $L_n^g$, and a Kac-Moody algebra for $(L_n^k, J_m)$. This decomposition of symmetry algebras leads to a decomposition of the Hilbert space into a product of sums of irreducible representations of the gravitational Virasoro and the Kac-Moody algebra, respectively,
	\begin{align}\label{factorization}
		\mathcal{H}_{tot}=\mathcal{H}_{g} \otimes \mathcal{H}_{k}  \,.
	\end{align}
	The density matrix of the system inherits this structure from Hilbert space
	\begin{align}
		\rho= \rho_{g} \otimes \rho_{k} \,.
	\end{align}
	This leads to a decoupling of the the charged moments in the two sectors,  \eqref{chargedMoments}, 
	\begin{eqnarray}\label{partition function}
		\mathcal{Z}_{n}(\mu)&=\Tr\left[(\rho_{\mathcal{A},g}^{n} \otimes \rho_{\mathcal{A},k}^{n}) e^{i\mu Q_{\mathcal{A}}}\right]
		&= \Tr\left[ \rho_{\mathcal{A},g}^{n} \right]		 \cdot \Tr\left[ \rho_{\mathcal{A},k}^{n} e^{i\mu Q_{\mathcal{A}}}\right] \,,
	\end{eqnarray}
	where the operator $e^{i\mu Q_{\mathcal{A}}}$ acts only on the $\u(1)_k$ sector. In particular, as shown in \cite{Zhao:2020qmn}, this nonlocal operator $e^{i\mu Q_{\mathcal{A}}}$ can be rewritten as the bilocal vertex operators
	\begin{align}
		e^{i\mu Q_{\mathcal{A}}}=V_{\mu}(v_1,\bar{v}_1)V_{-\mu}(v_2,\bar{v}_2)\,,
	\end{align}
	where $z=v_1$ and $z=v_2$ denotes the endpoints of the interval $\mathcal{A}$. The conformal weight and $U(1)$ charge of the vertex operator $V_{\mu}$ are given by \cite{Zhao:2020qmn}
	\begin{align}\label{eq: charge and conformal weight of vertex}
		\Delta_\mu=\bar{\Delta}_{\mu}=\frac{k}{4}\left(\frac{\mu}{2\pi}\right)^2\ , \quad \alpha_\mu=\bar{\alpha}_{\mu}=\frac{k}{2}\left(\frac{\mu}{2\pi}\right)\
	\end{align}
	The first term in \eqref{partition function} is exactly the $\nth$ power of the density matrix in the gravitational sector, which gives the ordinary R\'enyi entropy of the gravity side, while the second piece adds $\mathcal{O}(1)$ contributions to the ordinary R\'enyi entropy.
	
	When switching from the $n$-sheeted Riemann surface to the twist picture, the factorization property \eqref{factorization} implies that the twist operators from both sectors appear separately. The gravitational sector in \eqref{partition function} can be calculated using the corresponding twist operator $\sigma^{g}_n$. For the $U(1)$ sector in \eqref{partition function}, the corresponding twist operators live on the same Hilbert space as the vertex operators \cite{ginsparg1988applied} which allows to define the normal ordered product of both operators when inserted at the same points
	\begin{equation}\label{def charged twistor operator}
		\sigma_{\mu,n}=\normord{\sigma_n^k V_{\mu}' }
		\quad
		\text{with}
		\quad
		V_{\mu}'=\prod_{i=1}^n V^{i}_{\mu/n}\ .
	\end{equation}
	Here $\sigma^{k}_{n}$ denotes the twist operator in the $U(1)$ sector and $V^{i}$ is the vertex operator in the $i^{\text{th}}$ copy of the CFT. The flux $\mu/n$ appears in $V^{i}_{\mu/n}$ due to the fact that in the replica $n$-fold the total flux $\mu$ is distributed symmetrically across all $n$ sheets. From \eqref{eq: charge and conformal weight of vertex} and \eqref{def charged twistor operator}, the conformal dimension and charge of this charged twist operator $\sigma_{n,\mu}$ are then given by (cf. \cite{Zhao:2020qmn} for details)
	\begin{align}\label{charge conformal weight for charged twist operator}
		\Delta_{n,\mu}=\Delta_n^{(k)}+\frac{\Delta_{\mu}}{n} \quad \text{and} \quad \alpha_{n,\mu}=\alpha_{\mu}\,.
	\end{align}
	where $\Delta_n^{(k)}=\frac{c_k}{24}\left(n-\frac{1}{n}\right)$ denotes the conformal dimension of the twist operator $\sigma_{n}^{k}$ in the $U(1)$ sector.
	
	As a first check of the factorization \eqref{factorization}, we calculate the vacuum charged moments
	\begin{align}
		\mathcal{Z}_{n}(\mu)
		&=\langle V_{\mu}(v_1, \bar{v}_1)V_{-\mu}(v_2, \bar{v_2})\rangle_{n} \label{chargedMomVacuum}\\
		&= \langle \sigma^g_{n}(v_1, \bar{v}_1)\Tilde{\sigma}^g_{n}(v_2, \bar{v}_2)\rangle \times \langle \sigma_{n,\mu}(v_1, \bar{v}_1)\Tilde{\sigma}_{n,-\mu}(v_2, \bar{v}_2)\rangle \label{corrFactorization} \\
		&=\left|\frac{v_{1}-v_{2}}{\delta}\right|^{-\frac{c_g+c_k}{6}\left(n-\frac{1}{n}\right)-\frac{k}{n}\left(\frac{\mu}{2 \pi}\right)^{2}}\ ,\label{Z vacuum}
	\end{align}
	where in the second step we moved from the $n$-replica picture to the twist field picture, indicated by the replacement $\langle\cdots\rangle_n\to\langle\cdots\rangle$. This result coincides with the result derived in \cite{Zhao:2020qmn} after substituting $c=c_g+c_k$. The corresponding generating function for the vacuum state is then given by the $\mu$ dependent part of \eqref{Z vacuum}
	\begin{align}\label{f vacuum}
		f_{n}(\mu)=\left|\frac{v_1-v_2}{\delta}\right|^{-\frac{k}{n}\left(\frac{\mu}{2\pi}\right)^2}\ .
	\end{align}
	\subsection{Uncharged excited state} \label{sec:UnchargedExcitation}
	
	In this section we provide a calculation on the CFT side of the symmetry-resolved entanglement entropy in the excited state dual to the conical defect geometry. The corresponding gravity result was first calculated in \cite{Zhao:2020qmn} and is reviewed in Sec.~\ref{section 2.3}.
	
	Following the holographic dictionary in AdS${}_3$, deformations of the geometry away from the Poincaré patch are dual to insertions of {heavy} operators on the CFT side with conformal weights of $\mathcal{O}(c)$  \cite{RangamaniTakayanagiBook}. Due to factorization \eqref{factorization}, current primary operators cannot act on the gravitational Hilbert space, which leaves operators from the gravitational sector as the only candidates that can deform the geometry. Therefore, the nontrivial part of the calculations in this subsection happens almost completely the gravity sector. In this heavy operator background 
	, Rényi entropies and their holographic duals were already calculated in \cite{asplund2015holographic} using the monodromy method, which we follow closely. A useful review of the monodromy method is given in appendix C of \cite{fitzpatrick2014universality}.
	
	We start by calculating the charged moments \eqref{chargedMoments} in the presence of heavy primary operators $\psi(z)$ with conformal weights $\Delta_\psi$ and $\bar{\Delta}_\psi$, both assumed to be of $\cO(c)$. Taking the $\nth$ power of its density matrix $\rho_\cA(\psi)=\ket{\psi}\bra{\psi}$ instructs us to insert a copy of $\psi$ on each of the $n$ sheets of the Riemann surface, $\psi^{(n)}(z,\bar{z})=\prod_{i=1}^n \psi(z, \bar{z}; i^{\text{th}} \ \text{sheet})$. Hence one arrives at 
	\begin{align}\label{Excited Zn}
		\mathcal{Z}^{\Psi}_n(\mu)&=\left\langle \psi^{(n)}(0) \psi^{(n)}(\infty) V_{\mu}(v_1, \bar{v}_1) V_{-\mu}(v_2, \bar{v}_2)\right\rangle_{n} \notag\\
		&=\langle \Psi (0) \sigma^{g}_{n}\left(v_1, \bar{v}_1 \right) \tilde{\sigma}^{g}_{n}\left(v_2,\bar{v}_2\right) \Psi (\infty)\rangle \times \langle \sigma_{n,\mu}(v_1, \bar{v}_1)\Tilde{\sigma}_{n,-\mu}(v_2,\bar{v}_2)\rangle \notag\\
		&=\langle \Psi (0) \sigma_{n}^g\left(x, \bar{x} \right) \tilde{\sigma}_{n}^{g}\left(1\right) \Psi (\infty)\rangle \times \langle \sigma_{n,\mu}(x, \bar{x})\Tilde{\sigma}_{n,-\mu}(1)\rangle \notag\\
		&\equiv  \mathcal{Z}^{\Psi}_{n}(x, \bar{x};\alpha_{\Psi}) \times \langle \sigma_{n,\mu}(x, \bar{x})\Tilde{\sigma}_{n,-\mu}(1)\rangle\,.
	\end{align}
	In going to the second line of \eqref{Excited Zn}, we moved to the twist field picture and used the factorization property to split the charged moments into the two sectors $\cH_g$ and $\cH_k$. The operator $\Psi$, being the twist field picture pendant of $\psi^{(n)}$, is the product of all $n$ copies of the field $\psi$ and has conformal weight $\Delta_{\Psi}=n\Delta_{\psi}$. In the third step, we fixed the points $(v_1, \bar{v}_1) \rightarrow (x, \bar{x})$ and $(v_2, \bar{v}_2)\rightarrow 1$ for convenience. The quantity $\alpha_{\Psi}=\sqrt{1-24\Delta_{\psi}/c_g}$ appearing in the last step is defined for later convenience and the explicit dependence on $\alpha_{\Psi}$ is spelled out below. 
	
	In a holographic theory\footnote{The assumptions usually made when using the monodromy method are a sparse low lying spectrum and sub-exponential growth of the OPE coefficients with respect to the central charge \cite{Hartman:2013mia}, which also are the usual assumptions for a holographic CFT.} four point functions such as $\mathcal{Z}^{\Psi}_{n}(x, \bar{x};\alpha_{\Psi})$, where two operators have conformal weights $\Delta_2=\Delta_1=c \cdot \epsilon$ with $\epsilon\ll1$ and two operators have conformal weights $\Delta_3=\Delta_4 \sim \mathcal{O}(c)$, can be calculated perturbatively in $\epsilon$ using the monodromy method  \cite{fitzpatrick2014universality}. This procedure was used in \cite{asplund2015holographic} to calculate the Rényi entropy in the limit  $n \rightarrow 1$ from the four-point function $\mathcal{Z}^{\Psi}_{n}$ in \eqref{Excited Zn}. Following \cite{fitzpatrick2014universality}, as their conformal weights scale in the $n \rightarrow 1$ limit as $c_g(n-1)$,  we treat the twist operators perturbatively. The result reads
	\begin{align}
		\mathcal{Z}^{\Psi}_{n}(x, \bar{x};\alpha_{\Psi})
		&= \exp{\left(-2 \Delta_{n}\left(2 \log \lvert \frac{1-x^{\alpha_{\Psi}}}{\alpha_{\Psi}} \rvert + (1-\alpha_{\Psi}) \log |x|  \right) \right)}\nonumber \\
		&\equiv \exp{\left(-\Delta_{n}\,F(x, \bar{x},\alpha_{\Psi}) \right)}\,, \label{eq:CMG}
	\end{align}
	where we assumed a heavy operator of vanishing conformal spin,\footnote{Heavy operator insertions with finite conformal spin correspond to conically singular space-times on the AdS${}_3$ side. As detailed in \cite{Casals:2016odj}, the ones fulfilling the extremality bound $|M| \geq |J|$ ($M<0$) are healthy, while the rest admits closed time-like curves. We thank Ignacio Reyes for clarifications on this point.} $\Delta_{\psi}=\overline{\Delta}_{\psi}$. The absolute value comes from the presence of both holomorphic and antiholomorphic contributions. The next step is to calculate the two-point function
	\begin{align}\label{eq:chargepart}
		\langle \sigma_{n,\mu}(x, \bar{x})\Tilde{\sigma}_{n,-\mu}(1)\rangle
		&=\left|\frac{x-1}{\delta}\right|^{-\frac{c_k}{6}(n-\frac{1}{n})-\frac{k}{n}(\frac{\mu}{2\pi})^2}\,,
	\end{align}
	which is fixed by conformal symmetry. 
	Having factorized the charged moment, we can Fourier transform the charge part \eqref{eq:chargepart} according to \eqref{chargedMomentsFourier} and obtain $\mathcal{Z}_{n}(x,\bar{x},q)$. By virtue of \eqref{SRRenyi}, this yields the  symmetry-resolved Rényi entropies for $n-1=\epsilon\ll 1$ and, in the $n \rightarrow 1$ limit,  the symmetry-resolved entanglement entropy
	\begin{align}\label{eq:symmetry-resolved entanglement entropyresult1}
		S_1^{\Psi}(q)
		&= \frac{c_g}{6} F(x,\bar{x},\alpha_{\Psi}) + \frac{c_k}{3} \log{\left|\frac{x-1}{\delta}\right|} - \frac{1}{2} \log \left( \frac{k}{\pi} \log{\left|\frac{x-1}{\delta}\right|}  \right) +\mathcal{O}(1)\ .
	\end{align}
	The $F(x,\bar{x},\alpha_{\Psi})$ term encodes the entanglement entropy in the gravity sector, and thus the length of the Ryu-Takayanagi geodesic, as a function of the strength of the excitation given by $\alpha_\Psi$. The $k$-dependent part does not change compared to the vacuum background, indicating that the current is not influenced by this modification of the bulk geometry. This is a manifestation of the factorization of the gravity sector and the $\u(1)_k$ sector discussed in \secref{sec:factorization}. The term with the coefficient $\frac{c_k}{3}$ is the $\u(1)_k$ contribution to the ordinary entanglement entropy which, due to factorization \eqref{factorization},
	is also independent of the bulk geometry. Since $c_k$ is finite in the large $c$ limit, its contribution is outweighed by that of $c_g$ and may be neglected.
	
	In order to further interpret this result we first note that, working in radial quantization and Euclidean time, the spatial interval $\mathcal{A}$ lies on the unit circle in $z$-coordinates.  Mapping the complex $z$-plane to the $w$-cylinder $z=e^{-iw}$, with $w=\phi+it_E$, the transformation of the correlator $\mathcal{Z}(x,\bar{x};\alpha_{\Psi})$ yields $F(x,\bar{x};\alpha_{\Psi}) \rightarrow 2\log \left| \frac{2 \sin \left(\frac{\Delta \phi}{2}\alpha_{\Psi}\right)}{\delta \cdot \alpha_{\Psi}} \right|$, where $x=e^{i\Delta \phi}$.
	This function represents exactly a geodesic length in conical defect AdS$_3$ of defect angle $\alpha_{\Psi}=N^{-1}$. In these new coordinates, the symmetry-resolved entanglement entropy reads
	\begin{align}\label{symmetry-resolved entanglement entropyunchargedPrimary}
		S_1^{\Psi}(q)=\frac{c_g}{3} \log \left| \frac{2 \sin \left(\frac{\Delta \phi}{2}\alpha_{\Psi}\right)}{\delta \cdot \alpha_{\Psi}} \right| +\frac{c_k}{3}\log{\left|\frac{2\sin{\left(\Delta\phi/2\right)}}{\delta}\right|}- \frac{1}{2} \log \left( \frac{k}{\pi} \log{\left| \frac{2 \sin(\Delta\phi/2)}{\delta}\right|}  \right) \,,
	\end{align}
	which agrees with the result \eqref{eq:symmetry-resolved entanglement entropyConicalDefect} for the conical defect after identifying the ordinary  entanglement entropy $S$ in \eqref{eq:symmetry-resolved entanglement entropyConicalDefect} with the first two terms in \eqref{symmetry-resolved entanglement entropyunchargedPrimary}.The first term is the leading contribution in the large $c$ limit and is  reproduced by the Ryu-Takayanagi formula on the gravity side. Following \cite{belin2020bulk}, the second term arises on the gravity side as a quantum correction due to bulk entanglement. 
	\subsection{Charged excited state}\label{sec:CFTChargedVacuum}
	
	In the previous subsection we considered an excitation in the gravitational sector in the form of a heavy operator insertion which deforms the geometry. In this subsection we consider the situation where the background is additionally\footnote{In total, the operators we insert to generate the bulk charged conical defect are only heavy in the gravitational sector in the sense that their scaling dimensions $\Delta\sim c$. However, since the Kac-Moody level in the algebra \protect\eqref{symmalgebra} can be scaled to any value, no assumption on the value of the charge of these operators is invoked in our calculations.} excited by a $U(1)$ vertex operator, which only acts on the $U(1)$ sector and generates a background $U(1)$ current. Since the Kac-Moody sector has $c_k=1$, this operator is not heavy. Inside the bulk, these vertex operators can be interpreted as a Wilson line defect \cite{Zhao:2020qmn}. As already shown on the gravity side in Sec.~\ref{section 2}, this appearance of the bulk current will shifts the Gaussian distribution of charge fluctuations and does not change the symmetry-resolved entanglement compared to the vacuum case. Calculating the symmetry-resolved entanglement entropy on the CFT side provides a nontrivial test of this AdS${}_3$ prediction.
	
	The starting point of the calculation again are the charged moments. We denote the background vertex operators as $V_{\nu}(0)$ and $V_{-\nu}(\infty)$, which have opposite charge due to the global $U(1)$ symmetry. Hence, the charged moments can be written as
	\begin{align}\label{eq2}
		\mathcal{Z}^{\nu}_{n}(\mu)=\langle V^{(n)}_{\nu}(0) V_{\mu}(v_1,\bar{v_1}) V_{-\mu}(v_2,\bar{v}_2) V^{(n)}_{-\nu}(\infty)\rangle_n\ ,
	\end{align}
	with
	\begin{align}
		V^{(n)}_{\pm\nu}(z,\bar{z}):=\prod_{i=1}^{n}V_{\pm\nu}(z,\bar{z},i^{\text{th}}\ \text{sheet})\ .
	\end{align}
	The difficulty now is that the OPE between the charged twist fields $\sigma_{n,\mu}$ and vertex operators $V_{\nu}$ is unknown, hence we are not able to compute the charged moments \eqref{eq2} in the twist field picture as in the previous cases. However, the structure of the vertex operator correlation function on the complex plane is known to be
	\begin{align}\label{vertex algebra}
		\langle \prod_{i=1}^{k}V_{\alpha_i}(z_i,\bar{z}_i)\rangle = \prod_{i<j} \left|z_i-z_j\right|^{\frac{k}{4\pi^2}\alpha_i \alpha_j}\ .
	\end{align}
	Therefore, instead of using the twist field description, one can directly compute the charged moments \eqref{eq2} on the replica $n$-fold. For convenience, we consider the following function $F^{v}_{n}(\mu)$, which was first introduced in \cite{Capizzi:2020jed},
	\begin{align}
		F_n^{\nu}(\mu)=\frac{f^{\nu}_{n}(\mu)}{f_{n}(\mu)}=\frac{\langle V^{(n)}_{\nu}(0) V_{\mu}(v_1,\bar{v_1}) V_{-\mu}(v_2,\bar{v}_2) V^{(n)}_{-\nu}(\infty)\rangle_n}{\langle V^{(n)}_{\nu}(n) V^{(n)}_{-\nu}(\infty)\rangle_n \langle  V_{\mu}(v_1,\bar{v_1}) V_{\mu}(v_2,\bar{v}_2) \rangle_n} \label{eq:ExcitedGeneratingFunction}\ ,
	\end{align}
	Here $f^{\nu}_n(\mu)$ is the corresponding generating function for the excited state, given by
	\begin{align}
		f^{\nu}_n(\mu):=\frac{\mathcal{Z}^{\nu}_n(\mu)}{\mathcal{Z}^{\nu}_n(0)} =\frac{\langle V^{(n)}_{\nu}(0) V_{\mu}(v_1,\bar{v_1}) V_{-\mu}(v_2,\bar{v}_2) V^{(n)}_{-\nu}(\infty)\rangle_n}{\langle V^{(n)}_{\nu}(n) V^{(n)}_{-\nu}(\infty)\rangle_n} \,,
	\end{align}
	and $f_n(\mu)$ is the generating function for vacuum state given by \eqref{f vacuum}.

	The advantage of this function $F^{\nu}_{\mu}$ is that it transforms as a scalar function under conformal transformation, since the transformation factor in the denominator and the numerator of \eqref{eq:ExcitedGeneratingFunction} cancels out. Therefore, by uniformization map \eqref{uniformization}, one further obtains
	\begin{align}
		F^{\nu}_{n}(\mu)=\frac{\langle \prod_{m=1}^{n}V_{\nu}(\xi_{2m-1},\bar{\xi}_{2m-1}) V_{-\mu}(0) V_{\mu}(\infty) \prod_{m=1}^{n}V_{-\nu}(\xi_{2m},\bar{\xi}_{2m})\rangle_C}{\langle \prod_{m=1}^{n}V_{\nu}(\xi_{2m-1},\bar{\xi}_{2m-1}) \prod_{m=1}^{n}V_{-\nu}(\xi_{2m},\bar{\xi}_{2m})\rangle_C \langle  V_{-\mu}(0) V_{\mu}(\infty)\rangle_C}
	\end{align}
	with
	\begin{align}
		\xi_{2m-1}=\left(\frac{v_2}{v_1}\right)^{\frac{1}{n}}e^{\frac{2\pi i m }{n}}\ ,\quad \xi_{2m}=e^{\frac{2\pi i m }{n}}\ ,\quad m=1,2,\cdots,n\ .
	\end{align}
	Since $\xi$-plane is a complex plane, by the vertex operator algebra \eqref{vertex algebra}, it is obvious that the contribution to the function $F^{\nu}_{n}(\mu)$ in \eqref{eq:ExcitedGeneratingFunction} originates purely from the terms containing $\mu \nu$ in their exponent,
	\begin{align}
		F^{\nu}_n(\mu)&=\lim_{w\to\infty}\prod_{m=1}^{n}\left(\left|\frac{\xi_{2m-1}-0}{\xi_{2m}-0}\cdot\frac{\xi_{2m}-w}{\xi_{2m-1}-w}\right|^{-\frac{k}{4\pi^2}\mu\nu}\right)\nonumber\\
		&=\prod_{m=1}^{n}\left(\left|\frac{\xi_{2m-1}-0}{\xi_{2m}-0}\right|^{\frac{k}{4\pi^2}\mu\nu}\right)\nonumber\\
		&=\left|\frac{v_2}{v_1}\right|^{-\frac{k}{4\pi^2}\mu \nu}
	\end{align}
	Multiplying the vacuum generating function $f_{n}(\mu)$ in \eqref{f vacuum} yields
	\begin{align}\label{eq:3.26}
		f^{\nu}_{n}(\mu)=\left|\frac{v_2}{v_1}\right|^{-\frac{k}{4\pi^2}\mu\nu}\left|\frac{v_1-v_2}{\delta}\right|^{-\frac{k}{n}\left(\frac{\mu}{2\pi}\right)^2}=e^{i\mu Q_0}\left|\frac{v_1-v_2}{\delta}\right|^{-\frac{k}{n}\left(\frac{\mu}{2\pi}\right)^2}\ ,
	\end{align}
	where, by comparing with \eqref{generatingPP}, in the last equality we identified $ Q_0=-\frac{ik\nu}{4\pi^2}\log{\left|\frac{v_1}{v_2}\right|}$ as the background sub-region charge. $Q_0$ can be independently calculated in CFT as follows: Using  
	\begin{align}
		\langle J(z)\rangle_{\nu}=\frac{\langle J(z)V_{\nu}(0)V_{-\nu}(\infty)\rangle}{\langle V_{\nu}(0)V_{-\nu}(\infty)\rangle}=\frac{k}{4\pi}\frac{\nu}{z}\,,
	\end{align}
	the holomorphic part of the background sub-region charge defined in \eqref{sub-region charge def} is given by
	\begin{align}
		q_0=\int_{v_2}^{v_1}\frac{dz}{2\pi i}\langle J(z) \rangle_{\nu}=-\frac{ik\nu}{8\pi^2}\log{\left|\frac{v_1}{v_2}\right|}\,.
	\end{align}
	Furthermore, from the relation \eqref{eq: charge and conformal weight of vertex}, we have $\alpha_{\nu}=\bar{\alpha}_{\nu}$. Together with the analogous result for $\bar q_0$,  we find
	\begin{align}
		Q_0=q_0+\bar{q}_0=-\frac{ik\nu}{4\pi^2}\log{\left|\frac{v_1}{v_2}\right|}
	\end{align}
	We hence find that our field theory result \eqref{eq:3.26} is in complete agreement with the gravity result \eqref{generatingPP}, which was obtained by using the generating function method. As we have argued in \secref{section 2.3} already, the symmetry-resolved entanglement entropy does not change compared to the vacuum case. 

	\subsection{Two disjoint intervals}\label{sec:CFTtwoIntervals}
	We now turn to the case of two disjoint intervals $\cA_1$ and $\cA_2$ ($\cA_1\cap\cA_2=\varnothing$) comprising the entangling interval $\cA=\cA_1\cup\cA_2$.  Similarly to the previous case we consider the charged moments \eqref{chargedMoments}.
	In this case the treatment of the operator $e^{i\mu Q_{\mathcal{A}}}$ requires further explanation: Our goal is to write it again as a product of vertex operators. In order to do this, we need to split the charge as $Q_{\mathcal{A}}=Q_1 + Q_2$, where $Q_1$ and $Q_2$ are the charges in the two respective sub-regions. Since the two sub-region charges commute, $[Q_1, Q_2]=0$, the exponential $e^{i\mu Q_{\mathcal{A}}}$ straightforwardly splits into the product of two exponentials, 	
	\begin{align}
		e^{i \mu Q_{\mathcal{A}}}=e^{i\mu Q_{1}} e^{i\mu Q_{2}} \,.
	\end{align}
	This allows to represent the charged moments as
	\begin{align}
		\langle e^{i\mu Q_{1}} e^{i\mu Q_{2}} \rangle_{n}& = \langle V_{\mu}\left(z_{1}\right) V_{-\mu}\left(z_{2}\right) V_{\mu}\left(z_{3}\right) V_{-\mu}\left(z_{4}\right)\rangle_{n} \nonumber\\
		&=\langle \sigma_{n}^{g}(z_1)\tilde{\sigma}_{n}^{g}(z_2) \sigma_{n}(z_3)^{g}\tilde{\sigma}_{n}^{g}(z_4)\rangle \nonumber\\
		& \times \langle \sigma_{n,\mu}(z_1)\tilde{\sigma}_{n,-\mu}(z_2) \sigma_{n}(z_3,\mu)\tilde{\sigma}_{n,-\mu}(z_4)\rangle
	\end{align}
	As demonstrated in \appref{appendix B}, this four point function of vertex operators on the replica manifold can be obtained from integrating the Knizhnik-Zamolodchikov (KZ) equation \cite{KnizhnikZamolodchikov}. A simpler  way is to work in the twist field picture and use the generating function method. From \eqref{charge conformal weight for charged twist operator}, the current on the $i^{\text{th}}$ single sheet of the $n$-replica manifold corresponding to the $i^{\text{th}}$ copy of current in twist picture is then given by 
	\begin{align}
		\langle \hat{J}(z)\rangle_{n,\mu} 
		= \langle \hat{J}_{i}(z)\rangle_{\mu}
		= \frac{k}{2n}\left(\frac{\mu}{2\pi}\right)\left(\frac{1}{z-z_1}-\frac{1}{z-z_2}+\frac{1}{z-z_3}-\frac{1}{z-z_4}\right)\,,
	\end{align}
	where $\corr{\cdots}_{n,\mu}$ denotes the expectation value in the replica picture, while $\corr{\cdots}_\mu$ is in the twist field picture. The sub-region charge is obtained via integration over the sub-region
	\begin{align}
		\langle \hat{q}_{\mathcal{A}}\rangle_{n,\mu}
		& = \left(\int_{z_2+\delta}^{z_1-\delta}+\int_{z_4+\delta}^{z_3-\delta}\right) \frac{dz}{2\pi i}\ \langle J(z)\rangle_{n,\mu} \nonumber \\
		& = \frac{i k\mu}{4\pi^2 n}\log\left(\frac{(z_1-z_2)(z_3-z_4)(z_1-z_4)(z_2-z_3)}{\delta^2(z_1-z_3)(z_2-z_4)}\right)\,.
	\end{align}
	Combined with its antiholomorphic counterpart this yields the generating function \eqref{generatingfunction},
	\begin{align}\label{fn cft}
		f_{n}(\mu)=\left|\frac{(z_1-z_2)(z_3-z_4)(z_1-z_4)(z_2-z_3)}{\delta^2(z_1-z_3)(z_2-z_4)}\right|^{-\frac{k\mu^2}{4\pi^2 n}}\,.
	\end{align}
	Finally, the symmetry-resolved entanglement is given by
	\begin{align}\label{eq:Sqtwointervals}
		S(q)=S-\frac{1}{2}\log\left(\frac{k}{\pi}\log\left|\frac{(z_1-z_2)(z_3-z_4)(z_1-z_4)(z_2-z_3)}{\delta^2(z_1-z_3)(z_2-z_4)}\right|\right)\,.
	\end{align}
	Notice that the phase transition of the entanglement entropy occurs when the cross ratio $x=\left|\frac{(z_1-z_2)(z_3-z_4)}{(z_1-z_3)(z_2-z_4)}\right|$becomes $\frac{1}{2}$. For symmetry-resolved entanglement,  the $k$-dependent contribution from the $U(1)$ sector will not affect the phase transition. The interpretation for the gravity dual of the charged moments is, however, different in the two phases; see \secref{section 2.4} for a discussion on this point.
	
	\section{Discussion and Outlook}\label{sec:Outlook}

	In this paper we performed several nontrivial tests of the recently proposed holographic dual to symmetry-resolved entanglement entropy in {\adscft} \cite{Zhao:2020qmn} in the framework of $U(1)$ Chern-Simons-Einstein-Hilbert gravity. We consider the case of uncharged and charged conical defects, as well as of an entangling region consisting of two intervals. In all cases, we find agreement between the gravity and field theory calculation. An important ingredient in our computations is the generating function method for the calculation of the charged moments developed in \cite{Zhao:2020qmn}. 
 The leading large $c$ piece of the entanglement entropy in the ground state of a two-dimensional conformal symmetry can be reduced to the calculation of a two-point function of twist operators which are fixed by conformal symmetry. On the other hand, the cases we consider involve the calculation of four-point functions such as \eqref{Excited Zn} involving heavy operators in the gravitational sector, which are not fixed by conformal symmetry alone. Nevertheless, the correlators of charged vertex operators such as in \eqref{eq:ExcitedGeneratingFunction} are instead fixed by the structure of the U(1) Kac-Moody algebra. Hence calculations involving the charged vertex constitute a non-trivial test of the proposal, involving the symmetry structure on both sides of the correspondence. This is analogous to the free boson CFT, where all correlators are fixed by the U(1) symmetry, owing to the integrability of the model (c.f. e.g. chapter 9 of \cite{francesco2012conformal}).

	Our results for the symmetry-resolved entanglement entropy -- the excited states in the single interval case \eqref{eq:symmetry-resolved entanglement entropyConicalDefect} and the two interval case \eqref{Sqtwointervalsgravity} in the vacuum state -- contain several terms which are subleading in the large $c$ counting. The question of which terms in e.g. \eqref{eq:symmetry-resolved entanglement entropyresult1} are universal, i.e. terms that cannot be removed by a change of cutoff, seems to depend on whether they appear in other quantum information measures besides the symmetry-resolved entanglement entropy as well. For example, \cite{belin2020bulk} argued that the $-1/2 \log k$ term in the entanglement entropy stemming from $AdS{}_3$ bulk entanglement is non-universal, as it can be removed by a cutoff redefinition. However, one cannot use the same cutoff redefinition to remove the  $-1/2 \log ( k/\pi)$ term in \eqref{eq:symmetry-resolved entanglement entropyresult1} at the same time. A similar argument has been made by \cite{CardyTonni} for the appearance of the the boundary entropy in the entanglement entropy by comparing to Reny\'i entropies.\footnote{Of course, allowing for different cutoff definitions in different quantum information measures allows to shift away such terms. However, in experiments \cite{klich2006measuring,klich2009quantum,abanin2012measuring,islam2015measuring,Lukin256,brydges2019probing}, the cutoff is usually fixed by the experimental realization, and one is not free to choose it at will.} An unambigous way to confirm the universality of certain terms in the symmetry-resolved entanglement entropy\footnote{Or in any cutoff-dependent quantum information measure for that matter} would be to construct a cutoff-independent combination that picks out exactly the terms of interest, similar to the topological entanglement entropy in 2+1-dimensional systems constructed in \cite{KitaevPreskill,LevinWen2}. We will leave an investigation of such a construction for symmetry-resolved entanglement entropy for future work. 
	
	An important ingredient for our CFT calculations is the factorization property of the CFT Hilbert space into a large $c$ gravitational and a $\u(1)_k$ charged sector. This factorization seems to be intrinsic to the $\u(1)_k$ Kac-Moody symmetry considered here. From the bulk point of view, the reason is that the $U(1)$ Chern-Simons theory does not couple directly to the gravitational sector, but only contributes boundary photons (see \cite{belin2020bulk} for another instance where this topological property was employed). We expect a breakdown of factorization in any theory in which the bulk gauge field sector non-trivially backreacts onto the gravitational sector via additional charged matter fields. It will be interesting to further investigate whether and how factorization breaks down in 
	e.g. Einstein-Maxwell(-Chern-Simons) theories \cite{fujita2016holographic}, or in higher spin gravity \cite{toappear}. 
	
	To the best of our knowledge, we have reported the first result for a symmetry-resolved entanglement entropy in the two-interval case. In \secref{sec:CFTtwoIntervals} we worked in the twist picture. Because the OPE of charged twist operators is unknown, we were not able to compute the four-point function of charged twist operators, required for the charged moments, directly. To circumvent this problem, we used the generating function \cite{Zhao:2020qmn}, which is determined by the subregion charge, to extract the relevant contributions from the $U(1)$ sector of the theory. The attentive reader might worry whether the generating function method is valid when applied to $n$-folds of higher genera, since, at first sight, the current \eqref{fn cft} derived in the twist field picture does not contain topological information about the replica $n$-fold. For instance, the current does not feature the modular parameters or periodicities of the replica $n$-fold explicitely. Therefore in order to confirm our result from \secref{sec:CFTtwoIntervals}, we present an alternative derivation of the charged moments for the case of $n=2$ directly on the replica 2-fold in \appref{appendix B} using the Knizhnik-Zamolodchikov equation. This computation reveals how the modular properties of the torus, i.e. the replice 2-fold, enter in the charged moments. By showing that the current obtained in the twist picture is conformally related to the current on the torus, we prove that the results obtained in \secref{sec:CFTtwoIntervals} and \appref{appendix B} coincide.
	
	Even though we worked with a large $c$ CFT, we stress that we did not rely on vacuum block dominance in the two interval case to compute the second summand in \eqref{eq:Sqtwointervals}. It carries the symmetry resolution and derives entirely from the $U(1)$ symmetry of the theory. Therefore, it contains contributions from all conformal families, not just the vacuum family of the theory. This is a remarkable feature, as  vacuum block dominance makes most computations in large $c$ CFTs tractable in the first place. Here, this is traced back to the factorization of Hilbert space \eqref{factorization}.

	In all the cases we consider, we do not find a breakdown of equipartition of entanglement, i.e. the independence of the symmetry-resolved entanglement entropy of the charge of the subsector considered. This is due to the simple structure of the $JJ$ OPE. In a recent paper \cite{Calabrese:2021qdv}, the symmetry-resolved entanglement in Wess-Zumino-Witten models for compact Lie group $G$ is investigated. Without resorting to Fourier transformation in the chemical potential $\mu$, they find the breakdown of equipartition at constant order, while it persists at leading order. Given the abundance of models with equipartition at leading order in $c$, it is interesting to look for models which depart from this behavior. In all cases mentioned above the symmetry currents are spin-one primaries. This suggests to search within examples with higher spin currents, for instance higher spin gravity and its CFT with $W_3$ symmetry. As we will demonstrate in an upcoming article \cite{toappear}, we indeed find that equipartition of entanglement breaks down even at leading order in $c$ in holographic spin-3 gravity. The origin of this striking difference with the spin-one case is traced back to the complexity of the $W_3$ algebra.  
	
	In general, it is interesting to study larger non-abelian algebrae with or without supersymmetry. Besides the higher spin algebrae discussed above, one prominent example among top-down holographic models is the D1/D5 system, whose dual CFT is governed by the small $\cN=(4,4)$ superconformal algebra. This set of symmetries is particularly rich and it is possible to resolve with respect to different bosonic or fermionic subalgebrae of the $\cN=(4,4)$ SCFT. It will be interesting to apply and possibly generalize the tools employed here and in \cite{Zhao:2020qmn} to this case. We expect equipartition to be broken in this system. Moreover, considering top-down models should explicitely shed light on the range of applicability of our factorization assumption.
	
	An interesting future application of symmetry-resolved entanglement measures may be in the context of holography for two-sided black holes. Here, a breakdown of Hilbert space factorization \cite{harlow2016wormholes,harlow2018factorization}  between the two CFTs at the two asymptotic boundaries can be traced back to hidden gauge degrees of freedom in the bulk, which prevent e.g. the cutting of Wilson lines that connect the two asymptotic boundaries. Understanding how symmetry-resolved entanglement works in gauge theories in general might shed additional light on the mechanism of non-factorization of the CFT Hilbert in the context of AdS/CFT. Moreover, it is interesting to investigate symmetry-resolved entanglement in top-down theories, such as \cite{Eberhardt:2018ouy, Eberhardt:2019niq}, in particular since it provides insight into the applicability of factorization in the boundary CFT. Finally, our results pave the way to treat other quantum information measures such as mutual information, relative entropy, entwinement, or complexity, within the framework of symmetry resolution.

	\acknowledgments
	We thank Johanna Erdmenger, Marius Gerbershagen, Per Kraus, Ignacio Reyes and Henri Scheppach for useful discussions. 
	R.M., C.N., K.W. and S.Z. acknowledge support by the Deutsche Forschungsgemeinschaft (DFG, German Research Foundation) under Germany's Excellence Strategy through the W\"urzburg‐Dresden Cluster of Excellence on Complexity and Topology in Quantum Matter ‐ ct.qmat (EXC 2147, project‐id 390858490). The work of R.M. and C.N. was furthermore supported via project id 258499086 - SFB 1170 ’ToCoTronics’. S.Z. is financially supported  by the China Scholarship Council.
	

	\begin{appendix}
		
		
	\section{Kac-Moody current and Knizhnik-Zamolodchikov equation on the replica $2$-fold}\label{appendix B}

In this appendix, we consider the $n=2$ charged moments for the two disjoint intervals. Instead of using charged twist operators in the $\Z_2$ twisted picture, we compute the generating function directly on the corresponding replica $2$-fold, whose topology is that of a torus. We first introduce necessary geometric aspects of the replica $2$-fold, in particular, the conformal mapping from the replica $z$-coordinates to $w$-coordinates on the torus. Then, by solving the Knizhnik-Zamolodchikov equation \cite{KnizhnikZamolodchikov}, which arises from the Sugawara construction of the stress tensor for the $U(1)$ sector, we re-derive the generating function $f_2(\mu)$. Even though the resulting generating function takes a different form than \eqref{fn cft}, we prove it is in fact identical to the result \eqref{fn cft} of \secref{sec:CFTtwoIntervals}. Hence, we confirm the results in \secref{sec:CFTtwoIntervals} for the $n=2$ case.

We start by briefly introducing required geometric aspects of the replica $2$-fold for the case of two disjoint intervals. The replica $2$-fold is the $2$-fold branched covering of a complex plane, with two branch cuts. For convenience, we move the branch points delimiting the branch cuts from $(z_1,z_2,z_3,z_4)$ to the loci $(0,x,1,\infty)$. The map from the 2-fold with $z$-coordinates to the torus with $w$-coordinates can be obtained by integrating the holomorphic differential \cite{dixon1987conformal}
\begin{align}
    \text{d}w=\frac{\text { const }}{\left[\left(z-z_{\infty}\right)(z-1)(z-x)(z)\right]^{1 / 2}} \text{d}z
\end{align}
which defines an elliptic curve. For our purpose, it is more convenient to introduce the corresponding inverse mapping of the elliptic curve, given by 
\begin{align}\label{mapping torus}
    z(w)=\frac{\wp(w)-e_{1}}{e_{2}-e_{1}}, \quad x \equiv \frac{e_{3}-e_{1}}{e_{2}-e_{1}}\,,
\end{align}
where $e_1=\wp\left( \frac{1}{2}\right)$, $e_2=\wp\left( \frac{\tau}{2}\right)$ and $e_3=\wp\left( \frac{\tau+1}{2}\right)$ are the three half-periods of the Weierstrass elliptic function $\wp(w)$ and the point $z_{\infty}=z(0)$ is divergent. This map allows to calculate the charge in two different pictures and then compare the results.

To calculate the charged moments on the replica $2$-fold, we first recall the Sugawara energy-momentum tensor
\begin{align}
	\hat{T}_{J}(w)=\frac{1}{k}(\hat{J} \hat{J})(w)
	=\sum_{n \in \mathbb{Z}} w^{-n-2} L_n\ , \quad L_n=\frac{1}{k}\sum_m :J_m J_{n-m}:\ .
\end{align}
This provides two distinct ways of expressing the action of $L_{-1}$ on primary fields $\phi_i(w_i)$. Once as derivative $\p_i$ and once as 
\begin{align}
	(L_{-1} \phi_i)(w_i)=\frac{2}{k}(J_{-1}J_{0}\phi_i)(w_i)=\frac{2q_i}{k}(J_{-1} \phi_i)(w_i)
\end{align}
where $q_i$ is the  $U(1)$ charge of $\phi_i(w_i)$. Subtracting both expressions for $L_{-1}$, we obtain the null equation
\begin{align}
	(\partial_{w_i}-\frac{2q_i}{k}J_{-1})\phi_i(w_i)=0\,. \label{eq:KZMode}
\end{align}
This equation is understood as constraint on correlators. 

The Ward identity of a $U(1)$ current on a general Riemann surface was discussed in \cite{eguchi1987conformal}. On the torus, it takes the form
\begin{align}\label{Ward identity on torus}
	\langle \hat{J}(w) \phi_1(w_1)\ldots \phi_m(w_m) \rangle=-\sum_{k=1}^{m}q_k \tilde{G}(w_k,w)+\oint_{a}dw \langle J(w)\phi_1(w_1)\ldots \phi_m(w_m) \rangle\ ,
\end{align}
where $\tilde{G}(w,z)$ is the modified Green kernel for $\partial_{\bar{w}}$ operator on torus \cite{eguchi1987conformal}, given by
\begin{equation}
	\tilde{G}(w, z)=\zeta(w-z)-2 (w-z)\eta_{1}- 2\pi i \operatorname{Im}z/\operatorname{Im}\tau\ .
\end{equation}
where we used the shorthand $\eta_{1}=\zeta\left(\frac{1}{2}\right)$ and $\zeta(w)$ is the Weierstrass Zeta function, which is an odd function with double periods $1$ and $\tau$. 
The second term on the right hand side of \eqref{Ward identity on torus} represents the zero mode contribution for the current on the torus. Its integration is carried out over the contractible cycle of the corresponding solid torus, denoted by $a$. 

In our case, this zero mode contribution from the second term in \eqref{Ward identity on torus} vanishes.
This is easily seen in the twist field picture as follows. The $a$ cycle represents the circle around $(z_1, z_2)$ or $(z_1, z_4)$ depending on which channels we are working on. Since the total charge of the vertex operators within the circle always vanishes in both cases, the second term on the right hand side of \eqref{Ward identity on torus} vanishes. Furthermore, due to the global $U(1)$ symmetry, which implies $\sum_{k=1}^m q_k=0$, the non-analytic piece $\operatorname{Im}w/\operatorname{Im}\tau$ of the Green kernel $\tilde{G}(w_k,w)$ in \eqref{Ward identity on torus} cancels out. Hence, we have
\begin{align}\label{ward 2}
	\langle \hat{J}(w)\phi_1(w_1)\ldots\phi_m(w_m)\rangle=\sum_{k=1}^m q_k \left[\zeta(w-w_k)-2 (w-w_k)\eta_{1}\right]\langle\phi_1(w_1)\ldots\phi_m(w_m)\rangle \ .
\end{align}
Therefore, by \eqref{ward 2}, the action of $J_{-1}$ on the field $\phi_i(w_i)$ reads
\begin{align}\label{eq:KZCurrent} 
	\left\langle\phi_{1}\left(w_{1}\right) \ldots\left(J_{-1} \phi_{i}\right)\left(w_{i}\right) \ldots \phi_{m}\left(w_{m}\right)\right\rangle \nonumber 
	=& \oint_{w_i} \frac{d w}{2 \pi i} \frac{1}{w-w_{i}}\left\langle J(w) \phi_{1}\left(w_{1}\right) \ldots \phi_{m}\left(w_m\right)\right\rangle\nonumber \\
	=&  \sum_{\substack{k=1 \\ k \neq i}}^m  q_k [\zeta(w_i-w_k)-2(w_i-w_k)\eta_1]\left\langle \phi_1(w_1)  \ldots \phi_{m}\left(w_m\right)\right\rangle \ .
\end{align}
Combining the null equation \eqref{eq:KZMode} and the action of $J_{-1}$ in a correlator \eqref{eq:KZCurrent}, we obtain the Knizhnik-Zamolodchikov equation on the torus 
\begin{align}\label{kz eq}
	\left(\partial_{w_i}-\frac{2q_i}{k} \sum_{\substack{k=1 \\ k \neq i}}^m  q_k \, [\zeta(w_i-w_k)-2(w_i-w_k)\eta_1] \right) \left\langle \phi_{1}\left(w_{1}\right) \ldots \phi_{m}\left(w_m\right)\right\rangle=0\ ,
\end{align}
with $i=1,\ldots , m$. To solve these ordinary differential equations, the following relations are useful
\begin{align}
	\sigma(w)= \exp\left(\eta_1 w^2 \right) \frac{\theta_1(w,q)}{\theta_1'(0,q)}\ , \quad \zeta(w)=\frac{\sigma'(w)}{\sigma(w)}\ ,
\end{align}
where $\sigma(z)$ is the Weierstrass sigma function and $\theta_{1}(w,q)$ is the Jacobi theta function with the nome $q=\exp\left( 2 \pi i \tau \right)$.
The solution to \eqref{kz eq} is fixed up to some constant, given by
\begin{align}\label{kz correlator 1}
	\langle \phi_1(w_1)\ldots\phi_{m}(w_m)\rangle\propto\prod_{j<i}^n\left(\frac{\theta_1(w_i-w_j, q)}{\theta_1'(0,q)}\right)^{2q_iq_j/k}\ .
\end{align}

Coming back to the $n=2$ charged moments $\mathcal{Z}_{2}(\mu)$ for two disjoint intervals, by the mapping \eqref{mapping torus}, we can transform the charged moments to the four point function of vertex operators on the torus, i.e.
\begin{align}
	Z_2(\mu)=\left\langle V_{\mu}\left(\frac{1}{2}\right)V_{-\mu}\left(\frac{1+\tau}{2}\right)V_{\mu}\left(\frac{\tau}{2}\right)V_{-\mu}(0)\right\rangle
\end{align}
The corresponding generating function $f_{2}(\mu)$ is the $\mu$-dependent part of $\mathcal{Z}_2(\mu)$. Therefore, by \eqref{kz correlator 1} and \eqref{eq: charge and conformal weight of vertex}, it is found that
\begin{align}\label{f2 generating function}
	f_{2}(\mu)&=\left|\frac{\theta_{1}(-\frac{\tau}{2},q)\theta_{1}(\frac{\tau}{2},q)\theta_{1}(\frac{1}{2},q)^2}{\theta_{1}(\frac{1-\tau}{2},q)\theta_{1}(\frac{1+\tau}{2},q)\theta_{1}'(0,q)^2 \epsilon^2}\right|^{-\frac{k\mu^2}{4\pi^2}}\nonumber\\
	&=\left|\frac{\theta_1(\frac{\tau}{2},q)\theta_1(\frac{1}{2},q)}{\theta_1(\frac{1+\tau}{2},q)\theta_{1}'(0,q)\epsilon}\right|^{-\frac{k\mu^2}{2\pi^2}}\\
	&=\left|\frac{\theta_2(0,q)\theta_4(0,q)}{\theta_3(0,q)\theta_{1}'(0,q)\epsilon}\right|^{-\frac{k\mu^2}{2\pi^2}}\ ,
\end{align}
where the contribution from the anti-holomorphic part of the vertex operators is also included. The cut-off $\epsilon$ in the $w$-coordinates is introduced in order to make the generating function $f_{2}(\mu)$ dimensionless, since $\theta_1(w_i-w_j,q)/\theta_1'(0,q)$ has length dimension one.

To connect \eqref{f2 generating function} with the previous result \eqref{fn cft}, which was obtained by working in the twist field picture, we recall a known fact for the $n=2$ replica manifold, namely that the cross ratio $x=\frac{(z_1-z_2)(z_3-z_4)}{(z_1-z_3)(z_2-z_4)}=\frac{e_3-e_1}{e_2-e_1}$ can be expressed as
\begin{align}
	x=\left(\frac{\theta_4(0,q)}{\theta_3(0,q)}\right)^4\ .
\end{align}
Furthermore, by the Jacobi identity for the theta functions, i.e. $\theta_3(0,q)^4=\theta_2(0,q)^4+\theta_4(0,q)^4$, we have
\begin{align}
	1-x=\left(\frac{\theta_2(0,q)}{\theta_3(0,q)}\right)^4\ .
\end{align}
Using the above relation, it is evident that \eqref{fn cft} for $n=2$ and \eqref{f2 generating function} have similar structure. One way to show that \eqref{f2 generating function} exactly equals \eqref{fn cft} is to employ the relation between the two cut-offs $\delta$ and $\epsilon$, which is induced by the conformal mapping \eqref{mapping torus}. Here we choose a different way. The idea is as follows. Since the result \eqref{fn cft} is obtained by the subregion charge, we can first show that one can get the same generating function $f_{2}(\mu)$ from the subregion charge on the torus. Since the subregion charge is the integration of the current on the two intervals $\mathcal{A}$, it is conformally invariant. Therefore, by showing that the current on the torus and the current in the twist field picture are related by the conformal mapping \eqref{mapping torus}, we can prove the equivalence between \eqref{fn cft} and \eqref{f2 generating function} for the $n=2$ case. 

Now we want to reproduce the result \eqref{f2 generating function} by calculating the subregion charge on the torus. From \eqref{ward 2}, the current for the charged moments $\mathcal{Z}_2(\mu)$ on the torus reads
\begin{align}\label{current torus f2}
	J(w)&=\frac{\left\langle \hat{J}(w) V_{\mu}\left(\frac{1}{2}\right)V_{-\mu}\left(\frac{1+\tau}{2}\right)V_{\mu}\left(\frac{\tau}{2}\right)V_{-\mu}(0)\right\rangle}{\left\langle V_{\mu}\left(\frac{1}{2}\right)V_{-\mu}\left(\frac{1+\tau}{2}\right)V_{\mu}\left(\frac{\tau}{2}\right)V_{-\mu}(0)\right\rangle}\nonumber\\
	&=\sum_{k=1}^{4} q_k \left[\zeta(w-w_k)-2 (w-w_k)\eta_{1}\right]\nonumber\\
	&=\frac{k\mu}{4\pi}\frac{d}{dw}\log{\left(\frac{\theta_{1}(w-\frac{1}{2},q)\theta_1(w-\frac{\tau}{2},q)}{\theta_1(w,q)\theta_1(w-\frac{1+\tau}{2},q)}\right)}
\end{align}
In the last step we use the relation
\begin{align}
	\zeta(w)-2w\eta_1=\frac{d}{dw}\log{\theta_{1}(w,q)}\ .
\end{align}
Therefore, integrating the current \eqref{current torus f2} on the two intervals yields the holomorphic part of subregion charge on the torus,
\begin{align}\label{charge hol torus}
	q_{\mathcal{A}}=\left(\int_{(1+\tau)/2+\epsilon}^{1/2-\epsilon}+\int_{\epsilon}^{\tau/2-\epsilon}\right)\ \frac{dw}{2\pi i}\  J(w)
	&\approx\frac{i k\mu}{2\pi^2}\log{\left(\frac{\theta_1(\frac{\tau}{2},q)\theta_1(\frac{1}{2},q)}{\theta_1(\frac{1+\tau}{2},q)\theta_{1}'(0,q)\epsilon}\right)}\ ,
\end{align}
Here, since $\theta_1(0,q)=0$, we put the cut-off $\epsilon$ around each endpoint of the two intervals in order to regularize the divergence,  i.e. $\log{\theta_1(\epsilon,q)}\approx \log{\left(\theta_1'(0,q)\epsilon\right)}$. Combining \eqref{charge hol torus} with the anti-holomorphic charge yields the total subregion charge
\begin{align}
	Q_{\mathcal{A}}=q_{\mathcal{A}}+\bar{q}_{\mathcal{A}}=\frac{ik\mu}{\pi^2}\log{\left|\frac{\theta_1(\frac{\tau}{2},q)\theta_1(\frac{1}{2},q)}{\theta_1(\frac{1+\tau}{2},q)\theta_{1}'(0,q)\epsilon}\right|}=\frac{ik\mu}{\pi^2}\log{\left|\frac{\theta_4(0,q)\theta_2(0,q)}{\theta_3(0,q)\theta_{1}'(0,q)\epsilon}\right|}\ .
\end{align}
The corresponding generating function is then given by
\begin{align}
	f_2(\mu)=\exp{\left(i\int d\mu\  Q_{\mathcal{A}}\right)}=\left|\frac{\theta_4(0,q)\theta_2(0,q)}{\theta_3(0,q)\theta_{1}'(0,q)\epsilon}\right|^{-\frac{k\mu^2}{2\pi^2}}\ ,
\end{align}
which agrees with the result in \eqref{f2 generating function}.

The remaining task is to show that the subregion charge obtained via charged twist operators in the $\Z_2$ twisted picture is identical to the subregion charge on the torus. As we state before, due to the conformal invariance of the subregion charge, a sufficient condition for this equality between \eqref{fn cft} with $n=2$ and \eqref{f2 generating function} is that the currents in both cases are related by the conformal mapping \eqref{mapping torus}.  

For convenience, we send the endpoints of the two intervals $\mathcal{A}$ from $(z_1,z_2,z_3,z_4)$ to $(0,x,1,\infty)$. Then, by \eqref{charge conformal weight for charged twist operator}, the corresponding current in $\Z_2$ twisted picture is given by
\begin{align}\label{current twist f2}
	J(z)&=\frac{\left\langle\hat{J}(z)\sigma_{2,\mu}(0)\tilde{\sigma}_{2,-\mu}(x)\sigma_{2,\mu}(1)\tilde{\sigma}_{2,-\mu}(\infty)\right\rangle}{\left\langle\sigma_{2,\mu}(0)\tilde{\sigma}_{2,-\mu}(x)\sigma_{2,\mu}(1)\tilde{\sigma}_{2,-\mu}(\infty)\right\rangle}  \nonumber\\
	&=\frac{k\mu}{8\pi}\left(\frac{1}{z}-\frac{1}{z-x}+\frac{1}{z-1}\right)
\end{align}
The conformal transformation of $J(z)$ via the mapping \eqref{mapping torus} yields
\begin{align}\label{check diff}
	\left(\frac{dz(w)}{dw}\right)J(z)&=\frac{k\mu}{8\pi}\frac{d}{dw}\log{\left(\frac{(\wp(w)-e_1)(\wp(w)-e_2)}{(\wp(w)-e_3)}\right)}\nonumber\\
	&=\frac{k\mu}{4\pi}\frac{d}{dw}\log{\left(\frac{\theta_2(w,q)\theta_4(w,q)}{\theta_1(w,q)\theta_3(w,q)}\right)}\nonumber\\
	&=\frac{k\mu}{4\pi}\frac{d}{dw}\log{\left(\frac{\theta_{1}(w-\frac{1}{2},q)\theta_1(w-\frac{\tau}{2},q)}{\theta_1(w,q)\theta_1(w-\frac{1+\tau}{2},q)}\right)}\nonumber\\
	&=J(w)
\end{align}
where we used the following relation between the Weierstrass $\wp$-function and theta functions,
\begin{align}
	\wp(w)-e_1=\left(\frac{\pi\theta_4(0,q)\theta_3(0,q)\theta_2(w,q)}{\theta_{1}(w,q)}\right)^2\ ,\nonumber\\
	\wp(w)-e_2=\left(\frac{\pi\theta_2(0,q)\theta_3(0,q)\theta_4(w,q)}{\theta_{1}(w,q)}\right)^2\ ,\nonumber\\
	\wp(w)-e_3=\left(\frac{\pi\theta_4(0,q)\theta_2(0,q)\theta_3(w,q)}{\theta_{1}(w,q)}\right)^2\ .
\end{align}
The result \eqref{check diff} shows that the current \eqref{current twist f2} obtained in the twist field picture is indeed related to the current \eqref{current torus f2} on the torus by the conformal transformation \eqref{mapping torus}. Therefore, we conclude that the subregion charges as well as the generating function obtained in two different ways are identical. This confirms that the calculation via the charged twist operators in \secref{sec:CFTtwoIntervals} is valid in the $n=2$ case. While we have not treated the case of $n>2$, all our considerations should hold even for larger $n$, in which case the corresponding replica $n$-fold has higher genus, i.e. $g=n-1$. 
	\end{appendix}
	\bibliographystyle{JHEP}
	\bibliography{../../library}
	
\end{document}